\documentclass[preprint,prd,tightenlines,nofootinbib,eqsecnum,superscriptaddress]{revtex4-2}

\pdfoutput=1

\usepackage{amsmath}    
\usepackage{amsfonts}   
\usepackage{amssymb}    
\usepackage{mathtools}  

\usepackage[T1]{fontenc}   
\usepackage{mathrsfs}      
\DeclareSymbolFontAlphabet{\mathrsfs}{rsfs}

\usepackage{graphicx}   

\usepackage{array}               
\usepackage{booktabs}            
\usepackage[table]{xcolor}       

\usepackage{orcidlink}  
\usepackage{hyperref}
\hypersetup{
    colorlinks = true,
    linktocpage,
    unicode,
    linkcolor  = blue,
    filecolor  = magenta,
    urlcolor   = cyan,
}

\definecolor{CiteColor}{rgb}{0, 0.5, 0}       
\definecolor{RefColor} {rgb}{0.55, 0, 0}      
\definecolor{URLColor} {rgb}{0, 0, 0.35}      

\hypersetup{citecolor=CiteColor, linkcolor=RefColor, urlcolor=URLColor}

\newcolumntype{C}{>{$}c<{$}}

\AtBeginDocument{%
  \heavyrulewidth  = .08em
  \lightrulewidth  = .05em
  \cmidrulewidth   = .03em
  \belowrulesep    = .65ex
  \belowbottomsep  = 0pt
  \aboverulesep    = .4ex
  \abovetopsep     = 0pt
  \cmidrulesep     = \doublerulesep
  \cmidrulekern    = .5em
  \defaultaddspace = .5em
}

\newcommand{\PBdone}[1]{\cellcolor{green!12}#1}   
\newcommand{\PBind} [1]{\cellcolor{blue!10}#1}    
\newcommand{\PBtodo}[1]{\cellcolor{orange!18}#1}  
\newcommand{\PBnz}  [1]{\cellcolor{gray!18}#1}    
\newcommand{\PBna}     {\cellcolor{black!4}\textendash}  

\newcommand{\quand}{\quad\text{and}\quad}   
\newcommand{\ud}   {\mathrm{d}}             
\newcommand{\ui}   {\mathrm{i}}             

\newcommand{\scL}{\mathscr{L}}   

\newcommand{\RR}{\mathbb{R}}
\newcommand{\CC}{\mathbb{C}}

\newcommand{\mcM}{\mathcal{M}}   
\newcommand{\mcN}{\mathcal{N}}   
\newcommand{\mcC}{\mathcal{C}}   
\newcommand{\mcP}{\mathcal{P}}   
\newcommand{\mcT}{\mathcal{T}}   
\newcommand{\mcR}{\mathcal{R}}   
\newcommand{\mcX}{\mathcal{X}}
\newcommand{\mcY}{\mathcal{Y}}
\newcommand{\mcZ}{\mathcal{Z}}
\newcommand{\mcL}{\mathcal{L}}   

\newcommand{\ph}       {\phantom{o}}          
\newcommand{\tildemu}  {\tilde{\mu}}          
\newcommand{\fX}       {\mathfrak{X}}         

\begin{document}

\title{Symplectic mechanics of relativistic spinning compact bodies. III.\texorpdfstring{\\}{}Quadratic‑in‑spin integrability in Type‑D Einstein spacetimes: persistence and breakdown
}

\author{Paul Ramond, \orcidlink{0000-0001-7123-0039}}\,\email{ramond@lpccaen.in2p3.fr}
\affiliation{Universit\'{e} de Caen Normandie, ENSICAEN, CNRS/IN2P3, LPC Caen UMR6534, F-14000 Caen, France}

\author{Soichiro Isoyama, \orcidlink{0000-0001-6247-2642}}\,\email{isoyama@yukawa.kyoto-u.ac.jp}
\affiliation{Department of Physics, National University of Singapore, 21 Lower Kent Ridge Rd, 119077 Singapore}

\author{Adrien Druart, \orcidlink{0000-0002-1679-5077}}\,\email{Adrien.Druart@ulb.be}
\affiliation{Universit\'{e} Libre de Bruxelles, International Solvay Institutes, CP 231, B-1050 Brussels, Belgium}


\begin{abstract}
We investigate the integrability of spinning compact body dynamics at quadratic order in spin in four-dimensional Einstein spacetimes admitting a non-degenerate Killing--Yano tensor.
Working within the Mathisson--Papapetrou--Tulczyjew--Dixon framework under the Tulczyjew--Dixon spin supplementary condition, we model the spin-induced quadrupole with a deformability parameter $\kappa$, where $\kappa=1$ corresponds to black holes. The dynamics is formulated as a Hamiltonian system on a 10-dimensional physical phase space obtained by Dirac--Bergmann reduction. For $\kappa=1$, we establish Liouville--Arnold integrability at quadratic order in spin by constructing five independent, Poisson-commuting first integrals, including a generalization of the Carter constant and the R\"udiger constant to quadratic-in-spin order in Einstein spacetimes beyond Kerr. For $\kappa \neq 1$, the R\"udiger and Carter constants are no longer conserved; integrability does not persist at this order.
All our results are carried out in a covariant manner and numerically verified, and Kerr is recovered as a special case. These results show that integrability can extend beyond Kerr and beyond the linear-in-spin regime, while its breakdown for $\kappa \neq 1$ points to the spin-induced quadrupole as a decisive probe of compact body structure. 
\end{abstract}

\maketitle
\tableofcontents

\clearpage

\section*{Introduction and Summary}

\subsection{Scientific Context and Scope}

The motion of a (neutral) spinning compact body in the Kerr spacetime is integrable at linear order in the body's spin \cite{Ra.Iso.IntegO1.26} (see also, e.g.,~\cite{Apos.96,Kubiznak:2011ay,KuCa.12, KuLeLuSe.16,WitzHJ.19,Andersson:2025bhq,Witzany:2026eqc}). This integrability is underpinned by the Killing–Yano (KY) symmetry~\cite{Yano.52} of the Kerr geometry, which furnishes enough conserved quantities~\cite{Carter:1968ks,Rudiger.I.81, Rudiger.II.83} to render the dynamics predictable by quadrature (see, e.g., \cite{WiPio.23,Skoupy:2024jsi,Piovano:2024yks,SkouWitz.25,Piovano:2025aro,Abhinove:2026prep} for recent developments). 
Integrability is the entry point for efficient waveform-generation schemes developed for 
asymmetric mass binary inspirals~\cite{Chua:2020stf,Katz:2021yft,Speri:2023jte,Chapman-Bird:2025xtd,Honet:2025lmk}: it guarantees the existence of enough constants of motion and action–angle variables~\cite{Ra.Iso.PapII.24,Witzany:2024ttz}, on which adiabatic and post-adiabatic inspiral models are built \cite{HinFle.08, MiPo.21, PoWa.21, Mathews:2025nyb, Lewis:2025ydo}.

At quadratic order in spin, the body's internal structure enters through a spin-induced quadrupole moment, controlled by a deformability parameter $\kappa$. Black holes have $\kappa = 1$, as dictated by exact Kerr multipole relations \cite{Poisson:1997ha}; neutron stars have $\kappa \approx 2$--$14$ depending on the equation of state \cite{Laarakkers:1997hb}; and exotic compact objects such as boson stars can have $\kappa \approx 10$--$150$, with $\kappa < 0$ also possible for gravastars \cite{Cardoso:2019rvt}.
Geodesic integrability in Kerr has been known since 1968 by Carter \cite{Carter.68}. 
At linear order in spin, R\"udiger \cite{Rudiger.I.81, Rudiger.II.83} constructed the relevant KY-generated constants of motion. 
Integrability of spinning particle motion was established in the Schwarzschild spacetime in \cite{Apos.96,KuLeLuSe.16} and extended to Kerr in \cite{Kubiznak:2011ay,WiStLu.19,WitzHJ.19}, with further investigations beyond Kerr in \cite{KuCa.12,Andersson:2025bhq,Kim:2026pty,Witzany:2026eqc}. In Paper~I \cite{Ra.Iso.IntegO1.26}, we gave a covariant proof for all spacetimes admitting a non-degenerate KY tensor.
%
The quadratic-in-spin regime remains open. Two questions arise naturally: does integrability persist when structure--dependent, quadratic-in-spin effects are included, and if so, for which values of $\kappa$? Is the answer specific to Kerr, or does it extend to the broader class of spacetimes sharing its KY symmetry?

In the Kerr spacetime, important progress has been made at quadratic order in spin. Comp\`ere, Druart, and Vines \cite{ComDru.22,ComDruVin.23} constructed a quadratic-in-spin generalization of the Carter constant and the linear-in-spin R\"udiger constant, and showed that both are conserved for $\kappa = 1$ under the Mathisson--Papapetrou--Tulczyjew--Dixon (MPTD) equations \cite{Ma.37, Pa.51, Tu.59, Di.74, Ha.12} with the Tulczyjew--Dixon (TD) spin supplementary condition (SSC) \cite{Tu.59, Di.64}. Building on this, one of us (P.R.) proved Liouville integrability of the full quadratic-in-spin system in Kerr for $\kappa = 1$ using Hamiltonian methods \cite{Ra.CQG.24}. 
For $\kappa \neq 1$, no such constants were found, suggesting that integrability does not persist for material or exotic compact bodies.
At the root of the $\kappa=1$ case is the exact Kerr multipole relation $Q = -a^2 M$ \cite{Th.80}, which ensures a precise cancellation between dipole-squared and quadrupole
contributions \cite{Ra.CQG.24}; for $\kappa \neq 1$, microphysics breaks this relation. 

These results, however, were established in Kerr alone, using coordinate-based calculations. The present work asks whether they are consequences of a deeper geometric structure. Quadratic-in-spin constants of motion have also been explored recently in post-Minkowskian settings \cite{AkGo.25} and in a $\sqrt{\text{Kerr}}$ background on flat spacetime \cite{Arkani-Hamed:2019ymq,deFiVi.26,Kim:2026pty}; here we pursue a different direction, working in curved Einstein spacetimes admitting a KY tensor.
The MPTD framework describes the motion of an extended body with multipole moments in a curved spacetime; at quadrupolar order, the spin-induced quadrupole couples the body's internal structure to the background curvature \cite{Ha.12, Blanchet:2013haa, Di.15, Levi:2018nxp}. A choice of SSC is required to close the system. We adopt the TD SSC, which uniquely specifies the body's representative worldline. This choice is not merely a ``gauge'' fixing: the KY-generated invariants---generalizations of the R\"udiger and Carter constants to the spinning case---are conserved under the MPTD dynamics only when the TD SSC is imposed \cite{ComDruVin.23}, whereas the Killing-vector invariants associated with spacetime isometries are conserved regardless of the SSC \cite{Di.74, Ha.12}. In addition, the TD SSC plays a privileged role in the Hamiltonian formulation of the MPTD equations, see Sec.~II.D in \cite{Ra.Iso.IntegO1.26} for details. 
The Hamiltonian formulation of the MPTD dynamics has been pursued within various frameworks \cite{Souri.70, Kun.72, Ba.al.09, Ma.15,dAKuvHo.15,  ViKuStHi.16, WiStLu.19}; see also efforts in, e.g., \cite{Po.06, Porto:2016pyg, Levi:2018nxp}.
In Paper~I \cite{Ra.Iso.IntegO1.26}, we developed a covariant formulation with a non-degenerate symplectic structure on the physical phase space, and used it to prove linear-in-spin integrability in all spacetimes admitting a KY tensor, without assuming the Einstein field equations. Here we extend this framework to quadratic order in spin, which necessitates a stronger geometric assumption: we restrict to Einstein ($\Lambda$-vacuum) spacetimes admitting a KY tensor.

\subsection{Main results and Contributions}

This work extends the linear-in-spin analysis of Paper~I \cite{Ra.Iso.IntegO1.26} to quadratic order in the secondary's spin. The main results are:

\begin{itemize}
\item {Section~\ref{sec:Ham}}---a covariant Hamiltonian formulation for the MPTD equations under the TD SSC, consistent at quadratic order in spin. The framework incorporates both dipole-squared effects \textit{and} the spin-induced quadrupole \eqref{Jabcd} on a 10-dimensional physical phase space $\mathcal{P}$, obtained via Dirac--Bergmann reduction. The Hamiltonian is an autonomous constant of motion identified with the conserved effective mass $\tilde{\mu}$ of the secondary, which differs from the dynamical mass $\mu$ (the norm of the four-momentum) at quadratic order. The formulation is independent of the background spacetime.
\item {Section~\ref{sec:Carter-Rudiger}}---the construction of a quadratic-in-spin generalized Carter constant $Q$ that is conserved for $\kappa=1$ in any Einstein spacetime admitting a KY tensor. This extends the Kerr-specific result of \cite{ComDruVin.23} by recasting their construction in a coordinate-free, null bivector formalism adapted to the spacetimes with the KY symmetry. 
\item {Section~\ref{sec:PBint}}---a proof that the system is Liouville--Arnold integrable for $\kappa=1$ in Einstein spacetimes with KY symmetry. We exhibit five independent, Poisson-commuting first integrals on $\mathcal{P}$: the Hamiltonian $H$, two Killing-vector invariants $\Xi$ and $\mathfrak{X}$, the linear-in-spin R\"udiger constant $K$, and the quadratic generalized Carter constant $Q$. The proof proceeds via covariant Poisson-bracket computations using the null bivector decomposition of Sec.~\ref{sec:KY}. For $\kappa \neq 1$, $K$ and $Q$ are no longer conserved and integrability is lost \cite{ComDruVin.23}.
\end{itemize}

A companion Mathematica notebook \cite{MMA_IntegO2} provides explicit coordinate verification of all identities derived in this work in a Kerr--Newman--de Sitter background, using the \emph{RGTC} package~\cite{Bonanos:2003RGTC, Bonanos:RGTC_webpage}.

For $\kappa=1$, integrability at quadratic order in spin opens the door to constructing action--angle variables~\cite{Schm.02, WitzAA.22}, frequency maps~\cite{Dean.99}, Hamiltonian frequencies and resonances \cite{Schm.02, BrGeHiPRL.15, ZeLuWi.20, Lynch.al.24}, Hamilton-Jacobi equation \cite{Carter.68,WitzHJ.19}, perturbation (KAM-type) theory \cite{HinFle.08, Xue.20}, symplectic integrators \cite{Wu.al.21, Wang.al.21} and post-adiabatic waveform models that incorporate the secondary's quadrupole structure~\cite{Fa.al.12,St.15,RahBha.23,Timogiannis:2023pop,Shahzadi:2025ebj,RahShaPouMat.26}, generalizing existing geodesic and linear-in-spin developments for EMRI modelling \cite{HinFle.08, MiPo.21,Mathews:2025nyb}. 
For $\kappa \neq 1$, the absence of enough invariants suggests that non-integrable (and thus chaotic) dynamics may arise at quadratic order in spin for neutron stars and exotic compact objects, as has been explored at (nonlinear in spin) dipolar order and in non-Kerr spacetimes in, e.g.,  \cite{Suzuki:1996gm,Suzuki:1999si,Apostolatos:2009vu,LukWit.21,Destounis:2021mqv,Destounis:2021rko}, with implications for parameter inference with their gravitational--wave signals. 
In addition to the spin-induced quadrupole treated here, tidal interactions with the background curvature also arise at quadrupolar order. They typically break integrability, as we explored in a dedicated follow-up work \cite{Ra.NoCarter.26}.

\subsection{Organization of the paper}

This paper is organized as follows. Sec.~\ref{sec:quad-MPTD} sets up the MPTD dynamics at
quadrupolar order with a spin-induced quadrupole, fixes the TD SSC, and identifies the conserved mass $\tilde{\mu}$.
Section~\ref{sec:Ham} presents the covariant Hamiltonian formulation on a 14-dimensional phase space and reduces to the 10-dimensional physical phase space $\mathcal{P}$ via a Dirac bracket.
Section~\ref{sec:KY} develops the geometry of spacetimes with the KY symmetry using null bivector calculus.
Section~\ref{sec:Carter-Rudiger} constructs the generalized Carter constant at quadratic order in spin.
Section~\ref{sec:PBint} contains the integrability proof.

Our conventions are summarized in App.~\ref{app:ConvNot} and Table~\ref{Table}. App.~\ref{App:KdS} provides Kerr--de Sitter coordinate expressions used for consistency checks in the companion notebook \cite{MMA_IntegO2}. Technical identities are collected in App.~\ref{app:HOdge-LevC}--\ref{app:Fa(bc)d}.

\section{Equations of motion at quadratic order in spin}
\label{sec:quad-MPTD}

This section sets up the equations of the dynamics of the secondary object in a fixed but otherwise arbitrary background spacetime, neglecting self-field effects. We briefly review the covariant Dixon-Harte framework \cite{Di.74,Di.15,Ha.12,Ha.15} which models the secondary object as a particle endowed with linear momentum $p_a$, angular momentum (spin) $S^{ab}$ and multipoles moments. The evolution equations for the momenta are the classical Mathisson–Papapetrou–Tulczyjew–Dixon (MPTD) \eqref{EEgen}, which we truncate at quadrupolar order \eqref{FNquad}. We also state our choice of spin supplementary condition (SSC) \eqref{TDSSC} and establish a relation between the particle's four-velocity and its momenta \eqref{momvelTD}, and the existence of mass quantity \eqref{mutilde} conserved at quadratic order in spin. 

All calculations are performed consistently at quadratic order in spin, with higher orders implicitly neglected. We may use the shorthand $O(N)$ for terms with $N$ powers of the $S^{ab}$ (e.g. $O(2)$ for quadratic-in-spin). We also emphasize that we keep the background spacetime generic in this section and the next Sec. \ref{sec:Ham}, and specialize to Einstein geometries only later, starting in Sec.~\ref{sec:KY}.

\subsection{Dixon-Harte formalism for a multipolar particle}
\label{sec:DH-formalism}

\subsubsection{Multipolar expansion and MPTD equations}

The Dixon-Harte formalism \cite{Di.74,Di.15,Ha.12,Ha.15} consists in performing a multipolar expansion of a material source given by its conserved stress-energy tensor $T_{ab}$. The body is then modelled as a point particle endowed with multipole moments, and these moments are defined as integrals of $T_{ab}$ over worldline-adapted hypersurfaces. 
The first three multipole moments are the monopole $I^{ab}=I^{(ab)}$, the dipole $I^{abc}=I^{a(bc)}$, and the quadrupole $I^{abcd}=I^{(ab)cd}=I^{ab(cd)}$ such that $I^{(abc)d}=0$, which we rewrite in more usual forms as
\begin{equation}
\label{IspSJ}
    I^{ab} = p^{(a} v^{b)}\,, \qquad
    I^{abc} = S^{a(b} v^{c)}\,, \qquad
    I^{abcd} = \frac{4}{3}\, J^{(a|c|b)d}\,.
\end{equation}
Here, $v^a$ is the tangent to the worldline (which is not necessarily the four-velocity), and these relations serve as definitions of the
linear momentum $p^a$ and the spin tensor $S^{ab}=-S^{ba}$ of the particle, as tensors along its worldline $\scL$. 

The tensor $J^{abcd}$ is also called a quadrupole moment and is completely equivalent to $I^{abcd}$; we will refer to it simply as the
particle's \textit{quadrupole} from now on. It possesses the following independent algebraic symmetries:
\begin{equation}
\label{SymJabcd}
    J^{(ab)cd} =0, \quad J^{ab(cd)}=0 \quand
    J^{a[bcd]} = 0,
\end{equation}
which are identical to that of the Riemann curvature tensor. This implies that $J^{abcd}$ has at most 20 independent components.

The Dixon-Harte formalism is constructed such that the conservation equation $\nabla_a T^{ab} = 0$ for the body's stress-energy tensor holds if and only if the equations\footnote{Combining the algebraic symmetries of $R_{abcd}$ and the antisymmetry of $S^{bc}$, one has $ R_{abcd} S^{bc} = -\tfrac{1}{2} R_{adbc} S^{bc}$, which can explain different formulations of the MPTD equations throughout the literature.}
\begin{subequations} \label{EEgen}
\begin{align}
	\dot{p}_a  &= R_{abcd}\, S^{bc}\, v^d + F_a\,, 
	\label{EoM}
	\\
	\dot{S}^{ab} &= 2\, p^{[a} v^{b]} + N^{ab}\,, 
	\label{EoP}
\end{align}
\end{subequations}
are satisfied along the worldline $\scL$, where a dot symbolises $v^a\nabla_a$ (the covariant derivative along $\scL$), $R_{abcd}$ is the background spacetime's curvature tensor, and $(F_a,\,N^{ab})$ are the body-specific force and torque arising from coupling of the quadrupolar and higher-order multipoles to curvature. Owing to tradition \cite{Ma.15,Pa.51,Tu.57,Di.64}, we call equations \eqref{EEgen} the Mathisson–Papapetrou–Tulczyjew–Dixon (MPTD) equations. This is a set of 10 ordinary differential equations (ODEs) for $(p_a,S^{ab})$ along $\scL$.

Crucially, note that the MPTD equations do not provide an evolution equation for the quadrupole $J^{abcd}$. This is expected, already in non-relativistic mechanics: the evolution of higher multipole is not universal and determined by the microscopic structure of the body. 
To close the ODE system \eqref{EEgen}, one must supply a constitutive model for $J^{abcd}$ in terms of the momenta, the background geometry, and any external fields. 
%

\subsubsection{Scalar quantities}

We define three scalar fields along $\scL$ from the momenta $(p_a,S^{ab})$: 
\begin{equation}\label{eq:def-norms}
        \mu^2 := -p_a p^a, \quad 
        S_\circ^2 := \frac{1}{2}S_{ab}S^{ab} \quand
        S_\star^2 := \frac{1}{8}\varepsilon_{abcd}S^{ab}S^{cd},   
\end{equation}
where $\varepsilon_{abcd}$ is the Levi-Civita tensor. The quantities $\mu$ and $S_\circ$ are called the {dynamical mass} and the {spin norm}, respectively; $S_\star$ is a pseudo-scalar built from the spin. In general, they are \emph{not} conserved under the generic MPTD system \eqref{EEgen}; 
their time dependence depends on the choice of representative worldline for the body via $v^a$ and, crucially, on the multipolar force and torque in \eqref{EEgen}. 

In contrast, scalar quantities that remain unconditionally conserved along $\scL$ under \eqref{EEgen} arise when the background spacetime possesses Killing vector fields.\footnote{These invariants are at the core of the \emph{generalized Killing field} approach to the framework \cite{Ha.12,Ha.15}.} 
For any Killing vector $k^a$ of the background, one has
\begin{equation}\label{defFk}
    \mathcal{F}_k := p_a k^a + \frac{1}{2} \nabla_a k_b S^{ab} \quad \Longrightarrow \quad \dot{\mathcal{F}}_k =  0,
\end{equation}
independently of the choice of the force and torque in \eqref{EEgen}. This is a non-trivial result that is \emph{not} derivable from the MPTD equations alone, and draws upon the details of the Dixon-Harte framework \cite{Di.74,Ha.12,Ha.15}.

\subsection{MPTD equations at quadrupolar order with a spin-induced quadrupole}
\label{sec:fullmodel}

The Dixon-Harte formalism can be carried to arbitrary multipolar order, and the body's true stress-energy tensor $T^{ab}$ is in correspondence with the infinite set of multipole moments derived from it \cite{Di.74,Ha.15}. In this work, we truncate the expansion at \emph{quadrupolar} order: octupoles and higher order moments are neglected. 
Accordingly, we assume that the body under study is fully characterised by $(v^a,p_a,S^{ab},J^{abcd})$. 

At quadrupolar order, the force and torque in the MPTD equations \eqref{EEgen} are \cite{Di.74,Ha.12}
\begin{equation} \label{FNquad}
   F_{a}:= -\frac{1}{6} J^{bcde}\nabla_{a} R_{bcde} 
   \quand 
   N^{ab}:=-\frac{4}{3} J^{cde[a}R_{cde}^{\ph\ph\ph b]}.
\end{equation}
As in virtually all studies of quadratic-in-spin dynamics \cite{StPu.10,Bo.al.15,St.15,Ma.15,ViKuStHi.16,Chen.al.19,Timogiannis:2023pop,ComDruVin.23,RahBha.23,Ra.CQG.24,Shahzadi:2025ebj,RahShaPouMat.26}, we adopt a \emph{spin-induced} quadrupole in which $J^{abcd}$ is determined algebraically from $(p_a$, $S^{ab})$, the background curvature and an external parameter $\kappa$ that encodes the deformation of the extended body in response to its proper rotation: 
\begin{equation} \label{Jabcd}
    J^{abcd} = \frac{3\kappa}{\mu^3} p^{[a}\Theta^{b][c}p^{d]}\,,
    \quad \text{where} \quad 
    \Theta^{ab} := S^{ac}S_{c}^{\ph b}.
\end{equation}
As written, $J^{abcd}$ is quadratic in $S^{ab}$. Other nonlinear-in-spin terms arise from higher multipoles in the MPTD equations \cite{Ma.15,Vi.18}; to remain self-consistent we retain all quadratic-in-spin  terms and neglect higher multipoles in the subsequent analysis.
The parameter $\kappa$ in \eqref{Jabcd} encodes the deformation of the extended body in response to its proper rotation. It is defined such that $\kappa=1$ for an object deforming as an isolated Kerr black hole \cite{BiPo.09,ScheopnerVines.24}, and $\kappa > 1$ for other material compact object, larger values of $\kappa$ indicating ``stiffer'' equations of state \cite{LaPo.99,Bosh.al.12,Uchi.al.16}. Exotic compact object may even admit $\kappa < 0$ \cite{Cardoso:2019rvt}.

\subsection{The spin supplementary condition and its consequences}
\label{sec:TDSSC}

Since the angular momentum $S^{ab}$ is a bivector, one may perform a Hodge decomposition \cite{GouSR} on it with respect to the unit timelike vector
\begin{equation}\label{eq:def-hatp}
{\hat {p}}_a := p_a/\mu,
\end{equation}
which satisfies $\hat p^{a} \hat p_{a} = -1$. The Hodge decomposition effectively splits $S^{ab}$ into a spin vector $S^a$ and a mass dipole vector $C^a$, such that
\begin{equation}\label{zip!}
	S^{ab} = \varepsilon^{abcd} \hat{p}_c S_d + \frac{2}{\mu} C^{[a} \hat{p}^{b]} \quad \Longleftrightarrow 
	\quad \begin{cases}
		\, S^b := \frac{1}{2} \varepsilon^{abcd} \hat{p}_a S_{cd} \, , \\
		C^b := p_a S^{ab}  \, .
	\end{cases}
\end{equation}
The six degrees of freedom contained in $S^{ab}$ are thus encoded in the spacelike vectors $S^a$ and $C^a$, each with three independent components, as both are orthogonal to $p_a$. This is entirely analogous to the electric/magnetic decomposition of the Faraday tensor \cite{GouSR} (see also App.~II in Paper I \cite{Ra.Iso.IntegO1.26}). 

\subsubsection{Choice of spin supplementary conditions}

In this work, as in Paper I \cite{Ra.Iso.IntegO1.26} we adopt the ``Tulczyjew-Dixon'' spin supplementary condition (TD SSC) \cite{Tu.59,Di.64}, which sets the mass dipole $C^a$ defined above to zero : 
%
\begin{equation}\label{TDSSC}
\text{Enforcing the TD SSC} \quad \Leftrightarrow \quad 
C^b := p_a S^{ab} = 0 \quad \Leftrightarrow \quad 
S^{ab} = \varepsilon^{abcd} {\hat{p}}_c S_d \,. 
\end{equation}
This tensorial equation is equivalent to three \emph{algebraic} relations between the components $p_\alpha$ and $S^{\alpha\beta}$.\footnote{Only three independent equations, since $p_aS^{ab}=0$ is trivially true when projected into the $p_b$ direction.} The MPTD system \eqref{EEgen} under the TD SSC \eqref{TDSSC} is therefore an \emph{algebraic-differential} system of equations. While other SSCs are possible \cite{Be.67,Mad.69,Scha.I.79,Scha.II.79,CoNa.15,Co.al.12,Ha.15}, the TD SSC uniquely identifies the body's representative worldline and is particularly advantageous for our covariant Hamiltonian framework and the study of integrability; see Sec. II.D of Paper I \cite{Ra.Iso.IntegO1.26} for a more detailed rationale.

Using \eqref{zip!}, we derive the following decomposition of the tensor $\Theta^{ab}=S^{ac}S_{c}^{\ph b}$ involved in the expression for the quadrupole, cf. \eqref{Jabcd}:
\begin{equation}\label{expTheta}
    \Theta^{ab} = S^a S^b - S_{\circ}^2  h^{ab},\quad \text{with} \quad h^{ab}:=g^{ab}+\hat{p}^a \hat{p}^b,
\end{equation}
with $h^{ab}$ projecting orthogonally to $p_a$. Interestingly, note that $\Theta^{ab}$ is the projector onto 2-dimensional hyperplanes orthogonal to both $p^a$ and $S^a$, up to the factor $-S_\circ^2$.

\subsubsection{Momentum-velocity relation}

Combining the TD SSC \eqref{TDSSC} with the quadratic MPTD equations (i.e.,~\eqref{EEgen},~\eqref{FNquad} and~\eqref{Jabcd}) leads to a number of well-known results, recalled here for convenience. 

First, it implies an explicit relation for the particle's four-velocity $u^a$ --- whose components are
$u^\alpha := \ud x^\alpha / \ud\tau$
with $\tau$ the proper time along the worldline--- in terms solely of $(p_a,S^{ab})$ and the background geometry. Indeed, taking the covariant derivative of $p_aS^{ab}=0$ and using the MPTD equations \eqref{EEgen} yields the following expression\footnote{This result, as derived in \cite{GraHarWal.10} assumes, without loss of generality, that $v^a$ is normalized such that $v^\alpha p_\alpha = \mu$ for the sake of presenting streamlined formulae. We follow this convention as well.} \cite{EhRu.77,GraHarWal.10,ObuPue.11}
\begin{equation} \label{mo}
    \mu v^a 
    = 
    p^a + \frac{1}{2\mu^2}S^{ab} R_{bcde}p^c S^{de} - N^{ab}\hat{p}_b\,.
\end{equation}
This equation readily implies that $v^av_a=-1$ up to $O(3)$-terms: the tangent vector $v^a$ is equivalent to the four-velocity $u^a$ at $O(2)$. 
Using the spin-induced quadrupole \eqref{Jabcd} in the torque \eqref{FNquad}, one obtains the relation $p_a N^{ab} \;=\; \kappa\, h^{bc}\, R_{cade}\, \Theta^{de}\, \hat p^{a}$\footnote{
Detailed, this calculation is $p_a N^{ab} = \tfrac{4}{3} p_a J^{cde[a} R^{b]}_{\ph ecd} = \tfrac{2\kappa}{\mu^3} (p_a p^{[c} \Theta^{d] [e} p^{a]}R^b_{\ph ecd} - p_a p^{[c} \Theta^{d][e} p^{b]} R^a_{\ph ecd} )$. Using the symmetries and the orthogonalities reduces it to $ p_a N^{ab}= \kappa h^{ab} R_{acde} \Theta^{cd} \hat{p}^e$.} with $h^{ab}$ and $\Theta^{ab}$ defined in \eqref{expTheta}. Combining with \eqref{mo}, we obtain the momentum-velocity relation
\begin{equation}\label{momvelTD}
    u^a = \hat{p}^{a} + \frac{1}{2 \mu^3} S^{ab} R_{bcde} p^c S^{d e} + \frac{\kappa}{\mu^3} h^{ab} R_{bcde} \Theta^{cd}p^e,
\end{equation}
which agrees with (3.4) of \cite{Ma.15} or (3.5)-(3.6) from \cite{Hen.al.22}.
Notice that, once $p_a$ and $S^{ab}$ are known, equation \eqref{momvelTD} gives an ODE for the particle's worldline $\mathscr{L}$ since $u^\alpha (\tau) := \ud x^\alpha(\tau) / \ud\tau$ with $x^\alpha(\tau)$ denoting a parametrization of $\mathscr{L}$.

The momentum-velocity relation \eqref{momvelTD} implies that the familiar result $p^a\propto u^a$, which also holds in Newtonian mechanics and at geodesic and non-spinning orders, becomes false at $O(2)$. The last two terms in equation \eqref{momvelTD} are referred to as \emph{hidden momentum}, and are a relativistic feature. Notice also that even in the absence of quadrupole $(\kappa=0$ in \eqref{momvelTD}), there is still a quadratic-in-spin term, which is a dipole-squared effect, also absent in Newtonian mechanics.

Lastly, we mention that the two quadratic-in-spin terms on the RHS of \eqref{momvelTD} exhibit a remarkable simplification when $\kappa=1$ and the background is Einstein ($\Lambda$-vacuum). Indeed, using the decomposition of the spin tensors \eqref{zip!} and \eqref{expTheta}, the equality of both Riemann Hodge-duals for Einstein spacetimes \eqref{starR-Rstar}, and identity \eqref{epsIds}, we show that
\begin{equation} \label{p=muu}
   p^a = \mu u^a  + (1-\kappa)\, h^{a b} R_{bcde} S^c p^d S^e \quad \text{[in Einstein spacetimes]}.
\end{equation}
A few comments can be made on equation \eqref{p=muu}. First, in the case the ``canonical'' black hole binary (a Kerr secondary with $\kappa = 1$ in a Kerr primary), one obtains the remarkable result $p^a = \mu u^a$. This feature was already noticed in a Kerr background \cite{BiFaGe.15,ViKuStHi.16}. Our expression \eqref{p=muu} here provides a covariant proof of that result, while extending it to Einstein backgrounds. However, this proportionality between four-velocity and four-momentum fails generically in two ways: for non-Einstein backgrounds, with contributions from the Ricci tensor on the RHS of \eqref{p=muu}; or when the secondary does not possess the quadrupole structure of a test Kerr black hole, i.e., $\kappa\neq 1$.

\subsubsection{Conserved quantities under the TD SSC}

Combining the MPTD equations \eqref{EEgen} with the quadrupolar force and torque
\eqref{FNquad}, the momentum-velocity relation \eqref{momvelTD}, and the scalar
definitions \eqref{eq:def-norms}, one readily finds 
\begin{equation}
\label{SsunderSSC}
    \dot{S}_\circ = O(C^a), \qquad S_{\star} = O(C^a),
\end{equation}
where $O(C^a)$ denotes a term that vanishes \eqref{TDSSC} under the SSC. Upon applying the TD SSC \eqref{TDSSC}, this implies $S_\circ = \mathrm{const.}$
and $S_{\star} = 0$ at quadratic order in spin.

In contrast, the dynamical mass $\mu$ is not, in general, conserved at this order. Using \eqref{EoM}, \eqref{eq:def-norms}, \eqref{FNquad} and \eqref{momvelTD}, one finds $\dot{\mu} = (1/6)\, J^{abcd} \, \dot{R}_{abcd} + O(3)$, which needs not vanish. However, since the derivative of $J^{abcd}$ itself is at least $O(3)$ given its functional form \eqref{Jabcd}, the Leibniz rule clearly hints at a notion of approximately conserved mass. Indeed, the combination  
\begin{equation} \label{mutilde}
    \tilde{\mu} := \mu - \frac{1}{6}\, R_{abcd} J^{abcd} 
\end{equation}
is indeed conserved at quadratic order in spin \cite{DiI.70,Ma.15,Ha.23,ComDruVin.23}. 
Note that the mass $\tilde{\mu}$ is only conserved in this quadratic-in-spin model, i.e., with $J^{abcd}$ as given by \eqref{Jabcd}. For other quadrupole models, e.g., tidally induced, the numerical coefficient $1/6$ differs (see \cite{Chen.al.19,Ha.23,Ra.NoCarter.26}), or there may not even exist any notion of conserved mass for more generic quadrupole models.

\section{Hamiltonian formulation and phase space reduction} 
\label{sec:Ham}

This section provides the Hamiltonian backbone for our quadratic-in-spin $O(2)$ analysis. Building on the linear-in-spin results of Paper I, we incorporate the quadratic-in-spin effects (including spin-induced quadrupole \eqref{Jabcd}) into it and show that the quadrupolar MPTD dynamics with TD SSC (cf. Sec. \ref{sec:fullmodel}) are generated by the Hamiltonian 
\begin{equation}\label{def:HM-onshell}
H = - \frac{\tilde{\mu}^2}{2}, 
\end{equation}
where $\tilde{\mu}$ is the conserved mass \eqref{mutilde} at quadratic order, in the same way that $-\mu^2/2$ is the Hamiltonian for geodesics and linear-in-spin dynamics commonly used in the literature \cite{Schm.02,HinFle.08,WitzHJ.19,WiStLu.19,Ra.Iso.IntegO1.26,Ra.Iso.PapII.24}. 
The TD SSC, as an algebraic equation, cannot be a consequence of Hamilton's equations, which are ODEs. Rather, it is enforced onto phase space using the Dirac-Bergmann algorithm \cite{Bro.22}, as explained in Paper I in details (see Sec. III.C there), and reviewed in Sec.~\ref{10DPS} below. We subsequently obtain a 10-dimensional physical phase space, denoted $\mcP$, where trajectories described by Hamilton's equations corresponds to the solutions of the MPTD + TD SSC system (cf. Sec.~\ref{sec:HquadM} below). 
This reduced 10-dimensional Hamiltonian system is the platform for the integrability proof in Sec.~\ref{sec:PBint}, and exploits an important feature of the construction, presented in Sec.~\ref{sec:shortcut}.

\subsection{Poisson structure on the 14D phase space}
\label{subsec:gens}

To define our Hamiltonian system, we assume a given background metric and treat the spacetime coordinates and momenta components
\begin{equation} \label{PScoord}
(x^\alpha,p_\beta,S^{\gamma\delta})\in\RR^4\times \RR^4 \times \RR^6,
\end{equation}
as 14 coordinates spanning a 14-dimensional (14D) phase space, denoted $\mcM$. We will sometimes use the shorthand $(x,p,S)$ for these phase space coordinates. 

On the phase space $\mcM$, we consider the following Poisson brackets between all 14 phase space variables \eqref{PScoord}, already used in the literature \cite{Souri.70,Kun.72,dAKuvHo.15,Ra.CQG.24,Ra.Iso.IntegO1.26}
\begin{subequations}\label{PBs}
    \begin{align}
    &\{x^\alpha,p_\beta\} = \delta^\alpha_\beta \,, \\
    &\{p_\alpha,p_\beta\} =-\tfrac{1}{2}R_{\alpha\beta\gamma\delta}S^{\gamma\delta}\,, \label{PBpp} \\
    &\{p_\alpha,S^{\beta\gamma}\} =  2\Gamma^{[\gamma}_{\delta\alpha} S^{\beta]\delta} \,,  \\ 
    &\{S^{\alpha\beta},S^{\gamma\delta}\} = 2(g^{\alpha[\delta}S^{\gamma]\beta} 
    +g^{\beta[\gamma}S^{\delta]\alpha}) \,, \label{PBSS}
    \end{align}
\end{subequations}
where $\delta^\alpha_\beta$ is the 4D Kronecker symbol. The non-displayed Poisson brackets $\{x^\alpha,S^{\beta\gamma}\}$ and $\{x^\alpha,x^\beta\}$ identically vanish. The brackets \eqref{PBs} are anti-symmetric and satisfy the Jacobi identity (cf. Paper I \cite{Ra.Iso.IntegO1.26}). They exhibit key features of relativistic mechanics such as the local Lorentz algebra in the spin sector \eqref{PBSS} and spin-curvature coupling (sometimes called Papapetrou force) in \eqref{PBpp}. On the phase space submanifold where $S^{\alpha\beta}=0$ (no spin), the brackets reduce to those of four canonically-conjugated pairs $(x^\alpha,\,p_\beta)$ used for geodesics \cite{HinFle.08,Schm.02}. 

\subsection{The Hamiltonian and the evolution parameter}
\label{sec:HquadM}

At quadratic order in spin, we introduce the Hamiltonian $H$ as the following function of $(x^\alpha,p_\beta,S^{\gamma\delta})$ (a scalar field on $\mcM$) defined as follows\footnote{This Hamiltonian was first introduced in \cite{Ra.CQG.24}.}
\begin{equation} \label{HquadM}
    H(x,p,S) := 
    \frac{1}{2} g^{\alpha\beta} p_\alpha p_\beta 
    - \frac{1}{2} R_{\alpha \delta\beta \gamma} \hat{p}^\alpha S^{\beta\gamma} \hat{C}^\delta 
    + \frac{\kappa}{2} R_{\alpha \beta \gamma\delta} \hat{p}^\alpha \Theta^{\beta\gamma} \hat{p}^\delta,
\end{equation}
where a hat denotes normalization by the dynamical mass $\mu$ in \eqref{eq:def-norms} (see also~\eqref{eq:def-hatp}). Note that $C^\alpha$ does not vanish in general in $\eqref{HquadM}$: 
the quantity in $H(x,p,S)$ is a scalar function on $\mcM$, whereas the TD SSC holds only on a submanifold (this is the point of the subsequent reduction process).

The Hamiltonian \eqref{HquadM} reduces to the standard geodesic Hamiltonian \cite{HinFle.08,Schm.02} and the linear-in-spin formulations in \cite{Ra.Iso.IntegO1.26,Ra.Iso.PapII.24}. The second term (proportional to $C^\alpha$) is important. Though it vanishes under the TD SSC $C^\alpha=0$, once combined with Poisson brackets, it can produce terms that generally do not vanish under $C^\alpha=0$ (the two operations are not commuting). This was already pointed out in the pioneering work \cite{WiStLu.19} on pole-dipole hamiltonians. Yet, when $C^\alpha=0$, the right-hand side of \eqref{HquadM} is $-\tildemu^2/2$, with the conserved mass $\tildemu$ defined in \eqref{mutilde}. Indeed, the Hamiltonian itself reduces to a constant of motion under the TD SSC, as expected.\\

With the phase space variables \eqref{PScoord}, the Poisson brackets \eqref{PBs} and the Hamiltonian \eqref{HquadM}, the dynamics define a well-posed Hamiltonian system. To check equivalence with the MPTD equations at $O(2)$, one evaluates Hamilton’s equations using these brackets. For illustration, let us carry out the computation for the coordinate $x^\alpha$; this serves both to show how the formalism reproduces the MPTD dynamics in practice and to identify the natural evolution parameter associated with the Hamiltonian \eqref{HquadM}. 

Consider Hamilton's equation in its general form, which gives the evolution of any function $F := F(x,p,S)$ defined on the phase space $\mcM$:
\begin{equation} \label{HamEqM}
     \frac{\ud F}{\ud \lambda_H} := \{F,H\} = \frac{\partial F}{\partial x^\alpha} \{x^\alpha,H\} + \frac{\partial F}{\partial p_\alpha} \{p_\alpha,H\} + \frac{\partial F}{\partial S^{\alpha\beta}} \{S^{\alpha\beta},H\},
\end{equation}
where the first equality is the definition, and the second follows from the Leibniz rule on Poisson brackets. Upon inserting $F=x^\alpha$ and the Hamiltonian $H$ \eqref{HquadM} into \eqref{HamEqM}, and simplifying the result using the $\mcM$-brackets \eqref{PBs}, Hamilton's equation for $x^{\alpha}$ reads
\begin{eqnarray}
  \frac{\ud x^{\alpha}}{\ud \lambda_H} & = &  p^{\alpha} + \frac{1}{2 \mu^2} S^{\alpha \beta}  R_{\beta \gamma\delta \mu} p^\gamma S^{\delta \mu} + \frac{\kappa}{ \mu^4} p^{\alpha} R_{\beta \gamma\delta \mu} p^\beta \Theta^{\gamma\delta} p^\mu + \frac{\kappa}{\mu^2} g^{\alpha\beta} R_{\beta \gamma\delta \mu} \Theta^{\gamma\delta} p^\mu \notag \\
  & = & p^{\alpha} + \frac{1}{2 \mu^2} S^{\alpha \beta} R_{\beta \gamma\delta \mu} p^\gamma S^{\delta \mu} +\frac{\kappa}{\mu^2} h^{\alpha\beta} R_{\beta \gamma\delta \mu} \Theta^{\gamma\delta} p^\mu, \label{dxdlamH}
\end{eqnarray}
where we have: used the Leibniz rule applied to $\{x^\alpha,H\}$, combined the last two terms of the first line to introduce the projector $h^{ab}= g^{ab} + \hat{p}^a \hat{p}^b$, and made use of the following intermediary brackets
\begin{equation}
  \{ x^{\alpha},\, g^{\beta\gamma} p_\beta p_\gamma\} = 2p^{\alpha}, \quad 
 \{ x^{\alpha},\, \mu^{-2} \} = 2\mu^{-4}p^{\alpha} \quand 
  \{ x^{\alpha},\, C^\beta \} = S^{\alpha \beta}.
\end{equation}

We can now compare Hamilton's equation for $x^{\alpha}$ \eqref{dxdlamH} and the momentum-velocity relation \eqref{momvelTD}. The two results coincide if and only the Hamiltonian time $\lambda_H$ in \eqref{dxdlamH} satisfies 
\begin{equation} \label{lambdaH}
    \ud\lambda_H  = \frac{\ud \tau}{\mu}. 
\end{equation}
Since $\mu$ is not constant along phase-space trajectories, $\lambda_H$ is \emph{not} affine. It is not uncommon in relativistic mechanics to use non-affine parameters, such as the Mino time in Kerr spacetime \cite{Mi.03}. Yet, as given by \eqref{lambdaH}, $\lambda_H$ is the same parameter involved in the non-spinning (geodesic) \cite{Schm.02,HinFle.08} and linear-in-spin covariant formulation of the MPD equations \cite{WiStLu.19,Ra.Iso.PapII.24,Ra.Iso.IntegO1.26}.  Using \eqref{mutilde}, we can integrate \eqref{lambdaH} into
\begin{equation} \label{lambdaHint}
    \lambda_H= \frac{\tau}{\tilde{\mu}} - \frac{1}{6 \tilde{\mu}^2} \int R_{abcd} J^{abcd} \ud \tau, 
\end{equation} 
which shows that $\lambda_H$ is the (would-be affine) parameter $\tau/\tildemu$ with a (non-affine) $O(2)$-correction directly proportional to the quadrupole $J^{abcd}$. Note that we reach the same conclusion (that \eqref{lambdaHint} is the evolution parameter of the Hamiltonian) when comparing Hamilton's equations for the other phase space coordinates $(p_\alpha,\,S^{\alpha\beta})$ and comparing to the corresponding MPTD equations, namely \eqref{EoM} and \eqref{EoP}, respectively (the latter completed by \eqref{FNquad}-\eqref{Jabcd}). 
%

\subsection{Reduction to the 10D phase space via the TD SSC}
\label{10DPS}

At the level of the MPTD equations, the TD SSC~\eqref{TDSSC} imposes three independent algebraic constraints, and the four-velocity normalization $u^a u_a=-1$ imposes one. Since $(p_a,S^{ab},u^a)$ contain 14 components, these 4 constraints imply that a physical solution to the problem of motion can be described with $14-4=10$ independent variables. Hence, we expect that the phase space can be reduced to 10 dimensions, instead of 14 for now on $\mcM$. With $H$ and its evolution parameter fixed, we now explain how to impose properly the TD SSC in a Hamiltonian context, using the so-called Dirac brackets, and how the 10D physical phase space emerges. 

The account presented here is developed in greater details in Paper I \cite{Ra.Iso.IntegO1.26}, as well as in \cite{Ra.Iso.PapII.24} which is a direct application of the $\mcM\rightarrow\mcP$ reduction, to linear-in-spin dynamics in the Schwarzschild spacetime.

\subsubsection{Constraint surface and symplectic leaves}

On the 14D phase space $\mcM$, the variables $(x,p,S)$ are \emph{independent}; in particular they do not satisfy the TD SSC $C^\beta := p_\alpha S^{\alpha \beta} =0$ [cf.~Eq.~\eqref{TDSSC}]. Instead, the TD SSC defines a phase space submanifold in $\mathcal{M}$, denoted $\mcT$, defined as the set of points $(x^\alpha,p_\beta,S^{\gamma\delta})$ where the three equations $p_\alpha S^{\alpha\beta}=0$ hold. Consequently, $\mcT$ has dimension $14-3=11$, and, on it, the variables $(p_\alpha,S^{\beta\gamma})$ are \emph{no longer} independent.

In addition to this, the Poisson brackets \eqref{PBs} have a special feature that makes them \emph{degenerate}: there exist certain phase space functions on $\mcM$, called Casimirs, such that the Poisson bracket of \emph{any} function with a Casimir vanishes. This is a very particular type of constant of motion: typically, a constant of motion $I$ is one such that $\{I,H\}=0$. But here, a Casimir $\mathcal{S}$ is such that 
\begin{equation}
    \forall F \in \mathcal{C}^{\infty}(\mcM)\,,\quad \{F,\mathcal{S}\}=0.
\end{equation}

Casimirs are related to symmetry of the Poisson brackets themselves, whereas classical constants of motion $I$ are linked to symmetries of the Hamiltonian $H$. The Casimirs for our Poisson brackets \eqref{PBs} turn out to be the two spin invariants $(S_\circ,S_\star)$ defined in \eqref{eq:def-norms}. The higher-level symmetry related to them is that of the Lorentz group SO(1,3), which is hidden in the brackets \eqref{PBSS}.\footnote{These brackets, when projected onto an orthonormal tetrad (a class of tetrad fields invariant under Lorentz transforms), become $\{S^{AB},S^{CD}\}=2(\eta^{A[D}S^{C]B}+\eta^{B[C}S^{D]A})$, which is nothing but the structure generating of the Lorentz algebra so$(1,3)$ \cite{GouSR}.}

Lifting the degeneracies brought by these Casimirs is a simple matter: one simply projects onto the submanifold defined by their level sets. These submanifolds are called \emph{symplectic leaves} of $\mcM$, we denote them by $\mcN$. Since there are two Casimirs, $\text{dim}(\mcN)=14-2=12$. The submanifolds $\mcN$ are not degenerate any more, and are qualified as \emph{symplectic}, whence their name. \\

We work at the intersection $\mcT \cap \mcN$ of a fixed symplectic leaf under TD SSC.
Being in $\mcT$ implies that all solutions to Hamilton's equations satisfy the TD SSC \emph{by construction}, and being in $\mcN$ implies that the (projected) Poisson structure is free of degeneracies and we can work with usual Hamiltonian tools (existence of global canonical coordinates, Liouville's theorem, symplectic reduction, etc). Let us call this intersection $\mcP$:
\begin{equation} \label{defP}
    \mcP = \{(x,p,S)\in\mcM, \quad \text{s.t.} \quad p_\alpha S^{\alpha\beta}=0 \quand (S_\circ,S_\star)=\text{fixed}. \,\}
\end{equation}
i.e., $\mcP=\mcT\cap\mcN\subset\mcM$. Interestingly, since $C^a=0$ implies $S_\star=0$, only a 1-dimensional family of symplectic leaves $\mcN$ actually intersect $\mcT$: those with $(S_\circ,S_\star)\in\RR\times\{0\}$. Therefore, $\text{dim}(\mcP)=\text{dim}(\mcT)-1$, i.e., 
\begin{equation} \label{dimP}
\text{dim}(\mcP)=10.
\end{equation}
Importantly, we land on an even-dimensional submanifold (a prerequisite for symplectic-ness) and we end up with as many dimensions as fundamental degrees of freedom in the spinning MPTD + TD SSC framework, with $4+4+4-1-1=10$ encoded in $(p_a,S^a,u^a)$ and taking into account $p_a S^a=0$ and $u^au_a=-1$. 

\begin{figure}[t!]
    \begin{center}
    	\includegraphics[width=0.8\linewidth]{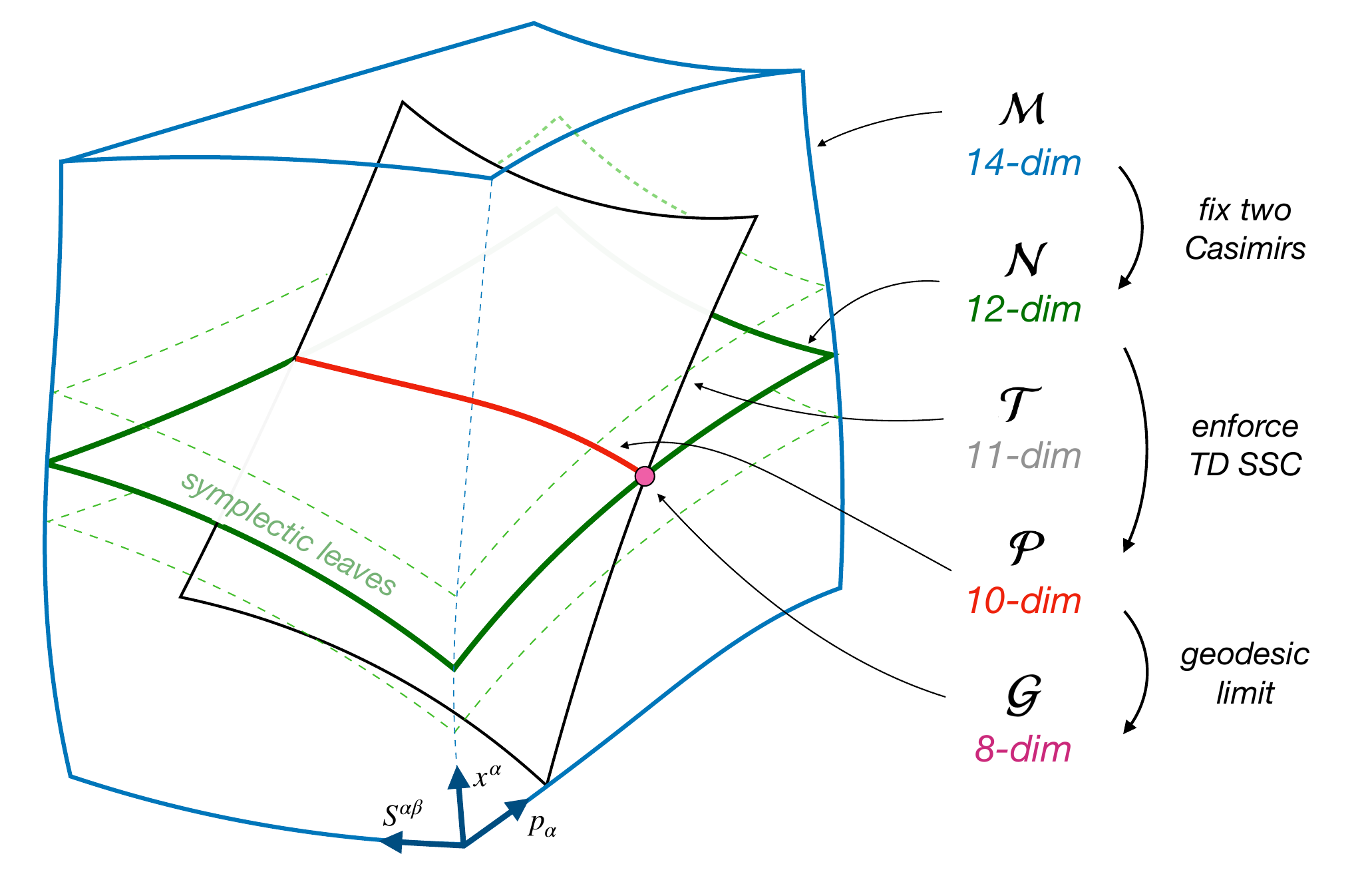}
        \caption{Different submanifolds are necessary to lift the degeneracies associated to the existence of Casimir invariants 
        $(\mcM\rightarrow\mcN)$ and correctly implement the TD SSC $(\mcN\rightarrow\mcT\rightarrow\mcP)$. The geodesic Hamiltonian \cite{Schm.02,HinFle.08} on the 8D phase space $(x^\alpha,\,p_\alpha)\in\mathcal{G}$ is simply obtained by setting $S^{\alpha\beta}=0$ which projects automatically onto the leaves $S_\circ=0$. }
        \label{fig:PS}
    \end{center}
\end{figure}

The above analysis is rather formal, and summarized geometrically in figure \ref{fig:PS}, but elementary when it comes to notions of Poisson geometry (the extension of Hamiltonian mechanics to phase space structures with degeneracies). 

\subsubsection{Stability under the Hamiltonian flow}

Importantly, the submanifold $\mcP$ is well-defined and, most crucially, \emph{stable} under the flow of $H$.\footnote{This stability is inherent to the framework construction, as the momentum-velocity relation (one of Hamilton's equations) already ``knows'' that the TD SSC is enforced.} Geometrically, this ensures that any solution to Hamilton's equations on $\mcM$ originating on $\mcP$ \emph{remains} within $\mcP$. This stability condition is expressed as
\begin{equation} \label{PstableflowH}
    \text{$\mcP$ is stable under the flow of $H$} 
    \quad \Longleftrightarrow \quad 
    \{H,C^\alpha\} = O(C^\alpha),
\end{equation}
which, considering \eqref{HamEqM}, exactly means the previous sentence. 

We verify~\eqref{PstableflowH} for the Hamiltonian \eqref{HquadM} by direct calculation. Using the Leibniz rule, we find
\begin{subequations} \label{PBCH}
    \begin{align}
    \{ C^\lambda, H \} & = - \tfrac{1}{2} \{ C^\lambda, \mu^2 \} + \{ C^e, \hat{C}^d \} R_{\alpha\beta\gamma\delta}\hat{p}^\alpha S^{\beta\gamma} + \tfrac{1}{2} \{ C^\lambda, R_{\alpha\beta\gamma\delta} \hat{p}^a \Theta^{\beta\gamma} \hat{p}^\delta \} + O (C^{\alpha})\\
    & = R_{\alpha \beta \gamma \delta} \left(S^{\beta \gamma} p^{\delta}  S^{\alpha \lambda} + \mu^{- 1} \{ C^\lambda, C^\delta \} \hat{p}^\alpha S^{\beta\gamma} + \tfrac{1}{2} \{ C^\lambda,\Theta^{\beta\gamma} \} \hat{p}^\alpha \hat{p}^\delta \right) + O(C^{\alpha})
    \end{align}
\end{subequations}
with $O(C^\alpha)$ denoting terms vanishing under the TD SSC $C^\alpha=0$. 
From Eq.~\eqref{PBs}, we derive the necessary auxiliary identities:
\begin{equation} \label{PBCC_PBCTheta}
    \{ C^{\alpha}, C^{\beta} \} = - \mu^2 S^{\alpha \beta} + O (C^{\alpha},3) 
  \quand 
  \{ C^{\alpha}, \Theta^{\beta \gamma} \} = 2 p^{[\beta} \Theta^{ \gamma] \alpha} + O (C^{\alpha}, 3),
\end{equation}
Substituting Eq.~\eqref{PBCC_PBCTheta} into Eq.~\eqref{PBCH} confirms that the condition \eqref{PstableflowH} holds for the Hamiltonian \eqref{HquadM}: the submanifold $\mcP$ is invariant under the Hamiltonian flow, allowing for a well-defined reduction to a Hamiltonian subsystem on $\mcP$. 

\subsubsection{Dirac bracket on $\mcP$}

To finally (and rigorously) pose the question of integrability of the MPTD + TD SSC, we need a last piece of machinery from Poisson geometry. Showing that a Hamiltonian system is integrable amounts to exhibiting enough constants of motion (called \emph{first integrals} in Hamiltonian mechanics) and performing a series of Poisson brackets between them to show their independence and involution. However, since we are only interested in the MPTD system under the TD SSC (otherwise even the Hamiltonian \eqref{HquadM} itself is not a constant), this means that we want to compute the Poisson brackets defined on the $10$-dimensional phase space $\mcP$, where things are non-degenerate, the TD SSC is enforced by construction and phase space curves are actual MPTD solution.

Therefore, we first need to find out what these Poisson brackets on $\mcP$ (or $\mcP$-brackets, for short) are. Once again, Poisson geometry has a compact and simple answer to this, which some readers may find as not so different from how tensors are projected onto spacetime hypersurfaces. Indeed, the $\mcP$-bracket can be computed from the $\mcM$-brackets \eqref{PBs}, and is a consequence of how Poisson structures are pulled back/pushed forward on invariant submanifolds \cite{Arn,Zambon.11,Burs.13,Derigl.22,Bro.22}. The formula reads
\begin{equation} \label{Diracall}
\{F,G\}^\mcP := \{F,G\}|_\mcP - \{F,C^\alpha\} (M^{-1})_{\alpha\beta} \{C^\beta,G\}|_\mcP \,, 
\end{equation}
for any two functions $(F,G)$ defined on $\mcM$, and
with $(M^{-1})_{\alpha\beta}$ denoting the matrix inverse of the matrix $M^{\alpha\beta}$ whose elements are $\{C^\alpha,C^\beta\}$. On the right-hand side of Eq.~\eqref{Diracall}, the $\mcM$-Poisson brackets are to be used, and then the symbol $|_\mcP$ means ``project onto $\mcP$'', i.e., ``simplify using $C^\alpha=0$''. Notice that the matrix $M$ has elements given explicitly by the first equality in \eqref{PBCC_PBCTheta}: its determinant does not vanish, $(M^{-1})_{\alpha\beta}$ exists, and the formula is well-posed. By construction, setting $G=C^\gamma$ into the $\mcP$-bracket formula \eqref{Diracall} implies that
\begin{equation} \label{CisCasimirofDB}
    \forall F\in\mathcal{C}^{\infty}(\mcM)\,,\quad \{F,C^\gamma\}^\mcP = 0 
\end{equation}
identically, as follows from $(M^{-1})_{\alpha\beta} \{C^\beta,C^\gamma\}=\delta_\alpha^\gamma$ by definition of the matrix $M$. Equation \eqref{CisCasimirofDB} simply tells us that $C^\alpha$ are Casimirs of the $\mcP$-bracket. This is how the TD SSC is enforced at the level of the Poisson structure.

\subsection{A practical bracket shortcut on $\mcP$} 
\label{sec:shortcut}

Lastly, we mention a crucial consequence from \eqref{Diracall} that will drastically simplify our forthcoming analysis. Consider a function $I\in\mcC^\infty(\mcM)$ such that $\{I,C^\alpha\}=O(C^\alpha)$. Then it follows from \eqref{Diracall} that 
\begin{equation} \label{Key}
   \forall F\in\mathcal{C}^{\infty}(\mcM)\,,\quad \{F,I\}^\mcP = \{F,I\}|_\mcP,
\end{equation}
since the second term in \eqref{Diracall} is $O(C^\alpha)$ from our assumption on $I$, and thus vanishes under $C^\alpha=0$. Consequently, the $\mcP$-bracket of any function with $I$ is simply the $\mcM$-bracket \eqref{PBs} whose result can be simplified using $C^\alpha=0$. An example of such a function is the Hamiltonian $H$ itself, which satisfies \eqref{PstableflowH}. 

The result \eqref{Key} is of considerable interest to us. We record this relation for future reference as the following proposition:
\begin{equation} \label{KeyResult}
   \{I,C^\alpha\}=O(C^\alpha) \quad \Longrightarrow \quad \forall F\in\mathcal{C}^{\infty}(\mcM)\,,\quad \{F,I\}^\mcP = \{F,I\}|_\mcP.
\end{equation}

Our integrability result, presented in Sec. \ref{sec:PBint}, will follow from the calculation of several $\mcP$-brackets. All of them will benefit from the result \eqref{KeyResult}, allowing us to never have to actually compute the second term in \eqref{Diracall}. This is because out of the five different first integrals (whose distinct pairs make up the $\binom{5}{2}=10$ $\mcP$-brackets), four satisfy the property of the function $I$ in \eqref{KeyResult}.  

\section{Killing–Yano Symmetry and Curvature Structure}
\label{sec:KY}

In this section, we review some general material about spacetimes with a Killing-Yano (KY) tensor in general relativity. Then, we establish the consequences of the existence of a KY using a formalism based on null decompositions presented in Sec.~\ref{sec:biv}. 
Our goal is to be self-contained and we only limit our results to the needs of subsequent sections. More details can be found in Paper I \cite{Ra.Iso.IntegO1.26}. Relevant material for KY tensors and their implication on spacetime geometry can be found in Refs. \cite{Carter.68,Ki.69,KiPhD.69,Coll.71,HughSomm.73,Coll.74,Hauser.I.75,Hauser.II.75,Coll.76,Steph.78,Rudiger.I.81,DiRu.I.81,DiRu.II.82,Rudiger.II.83,Wolf.98,JeziLuka.06,CookDray.09,AndBacTho.15,Frolov.17,Ha.20,LinSar.21,ComDru.22,ComDruVin.23,Ha.23}. 
%

\subsection{Killing-Yano tensor basics}
In addition to those arising from spacetime isometries, constants of motion also exist in spacetimes admitting symmetric Killing--St\"ackel (KS) tensors for geodesics \cite{Carter.68,WalkPen.70}, with the Carter constant in the Kerr spacetime as the most famous example. In general, however, a KS tensor is not sufficient to provide constants of motion beyond geodesic approximation. For example, constants of motion for the linear-in-spin dynamics rely instead on the existence of a KY tensor (which in particular implies the existence of a KS tensor) \cite{Rudiger.I.81,Rudiger.II.83,ComDru.22,Ra.Iso.IntegO1.26}. 

Accordingly, and as in Paper I \cite{Ra.Iso.IntegO1.26}, we search for constants of motion for quadratic-in-spin dynamics in spacetimes endowed with Killing-Yano symmetry. More precisely, we assume that the spacetime under study admits a KY tensor denoted by $f_{ab}$. It is defined as an antisymmetric tensor satisfying
\begin{equation} \label{deff}
    \nabla_{(a}f_{b)c}=0  \quand \mathfrak{Z}:=\tfrac{1}{4} f_{ab}^\star f^{ab} \neq 0,
\end{equation}
where, as in Paper I \cite{Ra.Iso.IntegO1.26}, the third condition ensures that $f_{ab}$ is non-degenerate \cite{DiRu.I.81,DiRu.II.82}. 

The existence of a KY tensor in a given spacetime strongly constraints its geometry, particularly its symmetries and curvature. 
With regard to symmetries, a KY automatically generates a symmetric KS tensor $K_{ab}$ and two commuting Killing vectors, which we denote by $\xi^a$ and $\eta^a$. They are defined by 
\begin{equation}\label{KSdef}
    K_{ab} := f_{ac}f^{c}_{\ph b}, \quad \xi_a := \frac{1}{3}\nabla^b f^\star_{ab} \quand \eta^a := K^{ab}\xi_b,
\end{equation}
and they satisfy the Killing and KS equations $\nabla_{(a}\xi_{b)}=0$, $\nabla_{(a}\eta_{b)}=0$ and $\nabla_{(a}K_{b)c}=0$. The identities that we use repeatedly are
\begin{subequations}
    \begin{align}
    f_{ac}^\star \,f^{c}_{\ph b}&=-\mathfrak{Z}\, g_{ab}, \label{ffstar} \\
    f_{ab} f_{cd} &=  -2f_{c[a} f_{b]d}  - \mathfrak{Z} \, \varepsilon_{abcd} ,  \label{fffeps} \\
    \nabla_a f_{bc} &=\varepsilon_{abcd}\xi^d, \label{defxi} \\
    \nabla_a f_{bc}^\star&=-2g_{a[b}\xi_{c]}. \label{dfstar}
    \end{align}
\end{subequations}

At the same time, the existence of a KY tensor has strong consequences regarding the curvature. 
By taking different covariant derivatives of the above identities and manipulating them using the Ricci identity, one obtains several integrability conditions for the Riemann tensor $R_{abcd}$. The ones that we will need are \cite{DiRu.I.81,DiRu.II.82,Ha.23}
\begin{subequations}
    \begin{align}
  R_{abe[c} f_{d]}^{\ph e} &= - R_{cde[a} f_{\ph b]}^e, \label{intcondcentral} \\
  R_{a b [c }^{\phantom{a c \phantom{di}} e} f^{\star}_{ d]
  e} &= - R_{c d [a }^{\phantom{a c \phantom{di}} e}
  f^{\star}_{ b] e}, \label{intcondcentral_star} \\
     \nabla_a \xi_b &= -\tfrac{1}{4}R_{ab}^{\ph\ph cd} f_{cd}^\star-\tfrac{1}{2}R_a^{\ph c} f^\star_{bc}. \label{Dxi} 
    \end{align}
\end{subequations}
A detailed exposition (including proofs) of the relations above is provided in Sec.~(IV) of Paper I \cite{Ra.Iso.IntegO1.26}. 

\subsection{Null decomposition} 
\label{sec:biv}

We now introduce the null bivector formalism, which will provide a natural framework to expose hidden symmetries inherited from KY tensors and connect them directly to conserved quantities in Sec.~\ref{sec:Carter-Rudiger}. At its core, the formalism draws on null tetrads and exploits the algebraic structure of spacetime geometries, not unlike the Newman–Penrose and Geroch-Hansen-Penrose frameworks \cite{NePe.62,NePe.66,GHP,PenRin1984,PenRin1986}). 
Below, we cover the essentials of this null bivector formalism, drawing on \cite{Ha.20,Ha.23} and referring to these references for conventions and notations.

\subsubsection{Bases for bivectors and symmetric tensors}

To begin, define a null tetrad consisting of four null vectors $(\ell^a, n^a, m^a, \bar{m}^a)$, where $\ell^a$ and $n^a$ are real, $m^a$ is complex, and $\bar{m}^a$ denotes the complex conjugate of $m^a$. These vectors satisfy the following orthogonality relations:
\begin{equation} \label{orthonlm}
    n^a\ell_a = -1 \quad \text{and} \quad m^a \bar{m}_a = 1,
\end{equation}
with all other inner products vanishing. The metric tensor can be written in terms of these null vectors as $g_{ab} = 2(\bar{m}_{(a}m_{b)}-\ell_{(a}n_{b)})$. Using the null tetrad, define three complex-valued two-forms as follows:
\begin{align} \label{defXYZ}
    X_{ab} = 2 \ell_{[a} m_{b]}, \quad Y_{ab} = 2 n_{[a} \bar{m}_{b]}, \quad
    Z_{ab} = 2 (\ell_{[a} n_{b]} - m_{[a} \bar{m}_{b]}),
\end{align}
as well as their complex conjugates, denoted $\bar{X}^{ab},\bar{Y}^{ab},\bar{Z}^{ab}$. These 6 bivectors form a basis of the six-dimensional vector space of complex-valued two-forms. They too obey orthogonality relations, which follow directly from equation \eqref{orthonlm}:
\begin{equation} \label{prodXYZ}
    Z_{ab}Z^{ab}=\bar{Z}_{ab}\bar{Z}^{ab}=-4,\quad X_{ab}Y^{ab}=\bar{X}_{ab}\bar{Y}^{ab}=-2,
\end{equation}
and all other inner products vanishing. By construction, the tensors $X^{ab},Y^{ab},Z^{ab}$ are self-dual, i.e., $X_{ab}^\star=\ui X_{ab}$, while their complex conjugate are anti self-dual, i.e. $\bar{X}_{ab}^\star=-\ui X_{ab}$, where a star denotes the Hodge dual: $X_{ab}^\star := \tfrac{1}{2} \varepsilon_{abcd}X^{cd}$ (see also App.~\ref{app:Hodge} for details).

It is also useful, for our purposes, to define the following tensors obtained by contracting the 2-forms with $\bar{Z}_{ab}$: 
\begin{equation} \label{symbasis1}
  \mcX_{a b}=\bar{Z}_a^{\ph  e} X_{e b}, \quad
  \mcY_{a b}=\bar{Z}_a^{\ph  e} Y_{e b}, \quad
  \mcZ_{a b}=\bar{Z}_a^{\ph  e} Z_{e b}.
\end{equation}

Importantly, $\mcX_{a b}, \mcY_{a b}, \mcZ_{a b}$ are all symmetric, and $\mcZ_{ab}$ is real-valued. These objects can be used to span the ten-dimensional vector space of symmetric real-valued tensors, by defining three more tensors such that
\begin{equation} \label{symbasis2}
    \mcL_{ab}=\ell_a\ell_b,\quad \mathcal{N}_{ab}=n_a n_b,\quad \mathcal{M}_{ab}=m_a m_b,
\end{equation}
of which $\mcL_{ab}$ and $\mathcal{N}_{ab}$ are real-valued and $\mathcal{M}_{ab}$ is complex-valued. We then have 10 real-valued symmetric tensors encoded in the following $4$ real-valued tensor $(g_{ab},\mathcal{Z}_{ab},\mcL_{ab},\mathcal{N}_{ab})$ and $3$ complex-valued tensors $(\mathcal{X}_{ab},\mathcal{Y}_{ab},\mathcal{M}_{ab})$. Aside from $g_{ab}$, all these tensors are trace-free. The non-vanishing scalar products between the elements are
\begin{equation} \label{orthosym}
    \mcL_{ab} \mcN^{ab} = \mcM_{ab} \bar{\mcM}^{ab} = 1 , \quad 
    \mcX_{ab}\mcY^{ab} = \bar{\mcX}_{ab}\bar{\mcY}^{ab}=2,\quad 
    \mcZ_{ab}\mcZ^{ab}=g_{ab}g^{ab}=4.
\end{equation}

Other identities satisfied by bivectors and symmetric tensors defined so far are given in App.~\ref{app:SS} and will be used throughout the paper. 

\subsection{Null decomposition of the Killing-Yano tensor}

There exists three types of tetrad transformations that leave the orthogonality conditions \eqref{orthonlm} unchanged \cite{Cha,Ha.15}. Among those, type-I transformations depend on a parameter $a \in \CC$, leave $\ell^a$ invariant and map $(n^c,m^d)$ to $(n^c + \bar{a}\, m^c + a\, \bar{m}^c + a \bar{a} \,\ell^c, m^d + a \ell^d)$, while type-II transformations depend on a parameter $b\in\CC$, leave $n^a$ invariant and map $(\ell^c,m^d)$ to $(\ell^c + \bar{b}\, m^c + b\, \bar{m}^c + b \bar{b}\, \ell^c,m^d + b\, n^d)$.
Through equation \eqref{defXYZ}, these transformations act on the complex bivectors $(X^{ab},Y^{ab},Z^{ab})$ as follows \cite{Ha.20}:
\begin{equation}
    (X_{ab},Y_{ab},Z_{ab}) \mapsto 
    \begin{cases}
    \, (X_{ab},Y_{ab}-\bar{a} Z_{ab}-\bar{a}^2 X_{ab},Z_{a b}+ 2\bar{a} X_{ab}), &\quad \text{[ Type-I ]} \\
    \, (X_{ab} + b Z_{ab} - b^2 Y_{ab},Y_{ab},Z_{a b} - 2 b Y_{ab}), &\quad \text{[ Type-II ]}
    \end{cases}
\end{equation}

Consider now the KY tensor $f_{ab}$, which is a real-valued bivecor. Its most general decomposition in the bivector basis reads $f_{a b} = 2 \text{Re} [ \alpha\, X_{a b} + \beta\, Y_{a b} + \gamma\, Z_{a b}]$, for some complex numbers $(\alpha, \beta, \gamma)$. Now, let us apply successively a type-I and a type-II transformation onto the null tetrad and choose the complex numbers $a$ and $b$ such that the new $X_{ab}$-- and $Y_{ab}$--components of $f_{ab}$ vanish. In the transformed tetrad (still denoted $(\ell^a, n^a, m^a, \bar{m}^a)$ as we will not transform away from it any more), $f_{ab}$ takes the form
\begin{equation} \label{fNP}
  f_{ab} = \text{Re} [C Z_{ab}],
\end{equation}
for a complex number $C$. This expression is valid in such transformed tetrad, which has been aligned with the principal null directions of the KY tensor \cite{Cha,Ha.15,Ha.20}. Notice that since $Z^{ab}$ is self-dual $(Z_{ab}^\star = \ui Z_{ab})$, taking the Hodge dual of \eqref{fNP} gives a simple expression for $f_{ab}^\star$, namely:
\begin{equation} \label{fstarNP}
  f_{ab}^\star = -\text{Im} [C Z_{ab}],
\end{equation}
where we used the property $\text{Re}[\ui z]=-\text{Im}[z]$ for any complex number $z\in\CC$. In fact, it will be useful to also have access to a complex-valued, self-dual version of the KY tensor, namely:
\begin{equation} \label{defPhi}
    \Phi_{ab} := f_{ab}-\ui f_{ab}^\star = C Z_{ab},
\end{equation}
of which $f_{ab}$ and $f_{ab}^\star$ are the real and (minus) the imaginary part. The tensor $\Phi_{ab}$ is a complex-conformal KY (CCKY) tensor, as it satisfies the CCKY equation (cf. \eqref{CCKY}). Regardless, in the tetrad adapted to $f_{ab}$, the latter has two independent components encoded in $C\in\CC$. The real and imaginary parts of $C$ are in relation with the two real-valued scalars built from the KY tensor: $f_{ab}^\star f^{ab}=4\mathfrak{Z}$ (already defined in \eqref{ffstar}) and $f_{ab}f^{ab}$ (which equals minus the trace of the KS tensor, cf. Eq. \eqref{KSdef}). The link between $C\in\CC$ and these two norms is
\begin{equation} \label{KYnorms}
    \text{Re}[C^2]=-\frac{1}{2}f_{ab}f^{ab} \quand \text{Im}[C^2]=2\mathfrak{Z}. 
\end{equation}

\subsection{Null decomposition of the curvature tensors}
\label{sec:geomKY}

Just like $f_{ab}$ is conveniently decomposed into the bivector basis according to \eqref{fNP}, one can perform a similar decomposition for curvature tensors. Let us start with the Weyl decomposition of the Riemann tensor $R_{abcd}$ in 4 dimensions \cite{Wald}:
\begin{equation} \label{Weyldec}
    R_{abcd} = C_{abcd} + g_{a[c} R_{d]b} - g_{b[c} R_{d]a} - \frac{1}{3} R\, g_{a[c} g_{d]b},
\end{equation}
wherein $C_{abcd}$ is the Weyl tensor, a trace-free tensor with the same symmetries as the Riemann tensor, and $R_{ab}:=R_{acbd}g^{cd}$ the Ricci tensor, whose trace $R:=R_{ab}g^{ab}$ is the Ricci scalar. The Riemann tensor has 20 independent components in four dimensions, that split into 10 components of $C_{abcd}$ and 10 components in $R_{ab}$, including $R$.

The decomposition of the Weyl tensor $C_{abcd}$ in the bivector basis \eqref{defXYZ} reads \cite{Cha,Ste}
\begin{align}\label{decompWeylNP}
  C_{abcd} =  2 \text{Re} \big[ &\Psi_0 Y_{ab} Y_{cd} + \Psi_1 (Y_{ab}
  Z_{cd} + Z_{ab} Y_{cd}) \nonumber \\
  & + \Psi_2  (Z_{ab} Z_{cd} - X_{ab} Y_{cd} - Y_{ab} X_{cd})\\
  & - \Psi_3 (X_{ab} Z_{cd} + Z_{ab} X_{cd}) + \Psi_4 X_{ab} \nonumber 
  X_{cd} \big],
\end{align}
where the five complex coefficients $(\Psi_0,\ldots,\Psi_4)$ are known as the Weyl scalars and can be obtained by contracting equation \eqref{decompWeylNP} with elements of the bivector basis and using the orthogonality relations \eqref{prodXYZ}. In particular \cite{Ha.20}:
\begin{equation}\label{defPsi2}
    \Psi_2=\frac{1}{16}C_{abcd}Z^{ab}Z^{cd}.
\end{equation}

Similarly, we can use the basis of symmetric tensors \eqref{symbasis1}--\eqref{symbasis2} to decompose the Ricci tensor $R_{ab}$ as
\begin{align} \label{decompRicciNP}
    R_{ab} = \Lambda g_{a b} + \phi_0 \mcL_{a b} + \phi_1 \mcN_{a b} + 2\phi_2 \mcZ_{a b} + 2 \text{Re}[\phi_3 \mcX_{a b} + \phi_4 \mcY_{a b} + \phi_5 \mcM_{a b}],
\end{align}
where the coefficients $(\Lambda,\phi_0,\ldots,\phi_5)$ are obtained by contracting equation \eqref{decompRicciNP} with the 10 symmetric tensors, and using the orthogonality relations \eqref{orthosym}. In particular:
\begin{equation} \label{defLambdaphi2}
    \Lambda=\frac{1}{4}R_{ab}g^{ab}\quand \phi_2 = \frac{1}{8}R_{ab}\mcZ^{ab}.
\end{equation}
The coefficients $(\Lambda,\phi_0,\ldots,\phi_5)$ are in a one-to-one relation with the Newman-Penrose coefficients traditionally used in the literature \cite{Cha,Ste}; we will not need their explicit forms.

\subsection{Constraints imposed by the KY tensor}

The decompositions \eqref{decompWeylNP} and \eqref{decompRicciNP} are the analogues of \eqref{fNP}--\eqref{defPhi} for the curvature tensors. They hold for any spacetime. However, in a spacetime endowed with a KY tensor, they simplify drastically. Indeed, let us follow the recipe: 
\begin{itemize}
    \item insert the decomposition \eqref{decompWeylNP}  for $C_{abcd}$ and \eqref{decompRicciNP} for $R_{ab}$ into the Weyl decomposition \eqref{Weyldec} for $R_{abcd}$, 
    \item insert the resulting expression for $R_{abcd}$ as well as the decompositions \eqref{fNP} for $f_{ab}$ and \eqref{fstarNP} for $f_{ab}^\star$ into the integrability conditions \eqref{intcondcentral}-\eqref{intcondcentral_star},
    \item contract the resulting expressions with appropriate combinations of the bivector basis \eqref{defXYZ} and symmetric basis \eqref{symbasis1}-\eqref{symbasis2} and simplify with the orthogonality \eqref{prodXYZ}-\eqref{orthosym}.
\end{itemize}

After this lengthy, but straightforward process, one finds that $(\Psi_0,\Psi_1,\Psi_3,\Psi_4)$ and $(\phi_0,\phi_1,\phi_3,\phi_4,\phi_5)$ all vanish.
Therefore, only $\Psi_2$, $\Lambda$ and $\phi_2$ remain in \eqref{decompWeylNP} and \eqref{decompRicciNP}, and the Weyl and Ricci tensors in the null tetrad adapted to $f_{ab}$ (where \eqref{fNP} holds) read
\begin{subequations}
    \begin{align} \label{finalCandCRicci}
    C_{abcd} &= 2\text{Re}[\Psi_2(Z_{ab}Z_{cd}-X_{ab}Y_{cd}-Y_{ab}X_{cd})], \\
    R_{ab} &= \Lambda g_{ab} + 2\phi_2 \mcZ_{ab}. \label{finalCandCRicci_bis}
\end{align}
\end{subequations}
These are the most general Weyl and Ricci tensors in a spacetime admitting a KY tensor. In particular, since only $\Psi_2$ remains, we have reproduced the classical result that spacetimes with a non-degenerate KY tensor must be of Petrov type-D\footnote{This follows from a series of works, with  \cite{Ki.69,KiPhD.69} for the original classification, \cite{Coll.74,Coll.76} for the vacuum case, and \cite{DiRu.I.81,DiRu.II.82} for the non-vacuum extension.}.

Interestingly, the Killing-Stäckel tensor has components in the same ``directions'' as the Ricci tensor \eqref{finalCandCRicci_bis}. Indeed, combining \eqref{KSdef}, \eqref{symbasis1} and \eqref{fNP}, we obtain 
\begin{equation} \label{decompKS}
    K_{a b} = \frac{1}{2} \text{Re} [C^2] g_{a b} + \frac{1}{2} |C|^2 \mcZ_{a b},
\end{equation}
which implies that any of $(g_{ab},\mathcal{Z}_{ab},K_{ab},R_{ab})$ can be expressed in terms of any  other two. 

Lastly, we give the following useful identity for a type-D Weyl tensor, obtained by expressing the combination $X_{ab}Y_{cd}-X_{cd}Y_{ab}$ in \eqref{finalCandCRicci} in terms of the metric and Levi-Civita tensor alone, using equation \eqref{zzxxyygg}. This reads
\begin{equation} \label{CZZggeps}
  C_{a b c d} = \text{Re} [\Psi_2 (3 Z_{ab} Z_{cd} + 2 g_{a [c} g_{d] b} -\mathrm{i} \varepsilon_{abcd})].
\end{equation}
%

\subsection{Summary: Einstein spacetimes with a KY tensor}

While \eqref{finalCandCRicci} are valid for \emph{any} spacetime endowed with a KY tensor, we will restrain our analysis to \emph{Einstein} spacetimes. These are defined by the Ricci tensor being proportional to the metric tensor:
\begin{equation} \label{DefEinstein}
    R_{ab}=\Lambda g_{ab}  \quad \Longleftrightarrow \quad R_{ab}-\frac{1}{2}Rg_{ab}+\Lambda g_{ab}=0,
\end{equation}
such that $g_{ab}$ is a solution of the general Einstein equations with zero stress-energy momentum tensor. Being an Einstein spacetime puts no constraints on the trace-free components of curvature, i.e., on the Weyl tensor. Equation \eqref{DefEinstein} together with \eqref{finalCandCRicci} implies that for Einstein spacetimes endowed admitting a KY tensor, $\phi_2=0$. 

To conclude, let us provide a final expression for the Riemann tensor $R_{abcd}$ of an Einstein spacetime endowed with a KY tensor, and its Hodge dual $R_{abcd}^\star:=\tfrac{1}{2}\varepsilon_{ab}^{\phantom{ab}ef}R_{efcd}$.\footnote{For Einstein spacetimes, both Hodge duals of the Riemann tensor coincide, as reviewed in App.~\ref{app:Hodge}} Combining \eqref{Weyldec}, \eqref{finalCandCRicci} and \eqref{DefEinstein} with $\phi_2=0$, 
we obtain
\begin{equation} \label{fullcurv}
    R_{abcd} = C_{abcd} + \frac{2\Lambda} {3}g_{a[c}g_{d]b}, \quand 
    R^\star_{abcd} = C^\star_{abcd} + \frac{\Lambda}{3}\varepsilon_{abcd},
\end{equation}
where the RHSs can be decomposed into the bivector basis \eqref{defXYZ} using \eqref{finalCandCRicci} and the identities in App.~\ref{ggXYZ_and_vepsXYZ};
we reproduce these identities below for the sake of clarity:
\begin{subequations}
    \begin{align}
        C_{abcd} &= 2\text{Re}[\Psi_2(Z_{ab}Z_{cd}-X_{ab}Y_{cd}-Y_{ab}X_{cd})], \label{CXYZ} \\ 
        g_{a[c}g_{d]b} &=- \frac{1}{2} \text{Re}[Z_{ab} Z_{cd} + 2 Y_{ab} X_{cd} + 2 X_{a b} Y_{cd}], \\
        \varepsilon_{abcd} &= \text{Im}[Z_{ab} Z_{cd} + 2 Y_{ab} X_{cd} + 2 X_{a b} Y_{cd}].
    \end{align}
\end{subequations}
These formulae, along with those for the KY tensor \eqref{fNP}, form the necessary ingredients used to compute the Poisson brackets ensuring our integrability result in Sec.~\ref{sec:PBint}.

Lastly, let us comment on the fact that the complex scalar $C$ determining the KY tensor (cf. \eqref{fNP}) is not independent from the curvature scalar $\Psi_2$, (though it is independent from the trace-part $\Lambda$). Indeed, using \eqref{fNP}, the KY equation $\nabla_{(a}f_{b)c}=0$ and the differential Bianchi identity $\nabla_{[a}R_{b]cde}=0$, one can establish that
\begin{align}\label{eq:C_Psi}
    \nabla_a(\Psi_2 C^3)=0\quad\Rightarrow\quad\Psi_2=\frac{\text{cst}}{C^3}.
\end{align}
This relation holds for any Einstein spacetime with a KY tensor. A detailed proof is provided in App.~ \ref{app:C_Psi}. For more general spacetimes, the differential relation is more complicated. In Kerr, an explicit calculation shows that $\Psi_2 C^3 = \ui M$ with the Kerr mass parameter $M$, which agrees with \eqref{eq:C_Psi}.

\section{Generalized Carter constant in Einstein spacetimes}
\label{sec:Carter-Rudiger}

In this section we construct a quadratic-in-spin generalization of the Carter constant beyond Kerr. Starting from the Kerr result of Comp\`{e}re--Druart--Vines \cite{ComDruVin.23}, we recast and extend their construction to all Einstein spacetimes admitting a KY tensor. The resulting scalar, which we call the \emph{Carter} constant (like its non-spinning counterpart), is built solely from the KY tensor and the particle's momenta and we show that it is conserved under the ${O}(2)$ MPTD dynamics with spin-induced quadrupole in the black-hole deformability case $\kappa=1$.\footnote{Throughout this section we restrict to the black-hole deformability case $\kappa = 1$, unless otherwise stated; the material-body deformability case $\kappa \neq 1$ does not admit a symmetry-generated quadratic invariant in general \cite{ComDruVin.23}.} 
Our approach draws on the Kerr-specific method of \cite{ComDru.22}, and exploits the KY-adapted null-bivector framework developed in Sec.~\ref{sec:KY}.

\subsection{Adapting reference \cite{ComDruVin.23} to our purposes}

We start with a very brief summary of the method and result presented in \cite{ComDruVin.23}. In a sentence, the authors generalized the result of Carter \cite{Carter.68} (resp. R\"udiger \cite{Rudiger.I.81,Rudiger.II.83}), who found a quadratic-in-momenta constant of motion $Q$ for geodesic (resp. linear-in-spin) dynamics in a Kerr spacetime. Their generalization amounted to adding a quadratic-in-spin correction to the previously found constant, in the form 
\begin{equation} \label{defQ}
    Q = K^{ab}p_a p_b + L_{abc}S^{ab}p^c + M_{abcd}S^{ab}S^{cd},
\end{equation}
where the first term on the RHS involves the KS tensor \eqref{KSdef}, the second term involves the tensor defined by \cite{Rudiger.I.81,ComDru.22}
\begin{equation} \label{defLabc}
    L_{abc}:= \frac{2}{3}\nabla_{[a}K_{b]c} + \frac{4}{3} \varepsilon_{abcd}\nabla^d \mathfrak{Z},
\end{equation}
and the last term in \eqref{defQ} involves a geometric tensor $M_{abcd}$ with the following symmetries (see Eq. (51) in \cite{ComDruVin.23}): 
\begin{equation} \label{symM}
    M_{(ab)cd}=0, \quad M_{ab(cd)}=0, \quand M_{abcd}=M_{cdab},
\end{equation}
as any difference from such symmetries would simply be cancelled when contracting with $S^{ab}S^{cd}$ to construct $Q$ in \eqref{defQ}. Applying the algorithm devised by R\"{u}diger \cite{Rudiger.I.81,Rudiger.II.83,ComDru.22}, the authors of \cite{ComDruVin.23} concluded by giving the following expression for the tensor $M_{abcd}$ (see their main result in Eq.~(133) of \cite{ComDruVin.23}), with $\xi^2:=\xi_a\xi^a$:
\begin{equation}\label{MCDV}
  M_{abcd} = - \xi_b g_{ac} \xi_d + \frac{1}{2} g_{ac} g_{db} \xi^2 + \frac{1}{2} f_a^{\ph e} R_{befd} f_{\ph c}^f  + \frac{1}{4} K_a^{\ph e} R_{ebcd}.
\end{equation}

The quantity $Q$ built following this procedure was shown to be a constant of motion for the MPTD + TD SSC equations at quadratic order in spin in the Kerr spacetime, provided that $\kappa=1$, i.e. that the test body exhibits the spin-induced quadrupole structure expected of a Kerr black hole \cite{ComDruVin.23}. Moreover, numerical evidences suggested that this constant of motion could not be build for other test bodies structures (i.e. in the $\kappa\neq 1$ case).

There is one point, crucial for our purposes, that was not imposed in \cite{ComDruVin.23}: the tensor $M_{abcd}$ provided in Eq.~\eqref{MCDV} does not satisfy the symmetries \eqref{symM}. This is inconsequential in \cite{ComDruVin.23}, since the contraction with $S^{ab}S^{cd}$ in \eqref{defQ} bypasses this lack of symmetries.  
However, the discrepancy between the stated symmetry \eqref{symM} and the explicit expression \eqref{MCDV} matters in our analysis, since $M_{abcd}$ will appear uncontracted from $S^{ab}S^{cd}$. We resolve this by making the following substitution
in formula \eqref{defQ} :
\begin{equation}\label{recipeMtildeM}
    M_{abcd} \longrightarrow \tilde{M}_{abcd} := \frac{1}{2} \bigl(M_{[ab][cd]} + M_{[cd][ab]}\bigr).
\end{equation}
This automatically imposes the symmetries \eqref{symM} while also ensuring that $\tilde{M}_{abcd}S^{ab}S^{cd}=M_{abcd}S^{ab}S^{cd}$ (even though $M_{abcd}\neq \tilde{M}_{abcd}$). In this fashion, the main result of \cite{ComDruVin.23} remains preserved, while the formula for the expression of $M_{abcd}$ has the symmetries claimed in \cite{ComDruVin.23}. Applying the recipe \eqref{recipeMtildeM} to the expression \eqref{MCDV} leads to the following \emph{definition} for the quantity $\tilde{M}_{abcd}$:
\begin{equation} \label{Mtilde}
  \tilde{M}_{abcd} := \xi_{[c} g_{d][a} \xi_{b]} 
  + \frac{1}{2}g_{c[a} g_{b]d}\xi^2 
  - \frac{1}{2} f_{[a}^{\ph e} R_{b]ef[c} f_{\ph d]}^{f} 
  + \frac{1}{8}( R_{cd[a}^{\phantom{abci}e} K_{b]e} + R_{ab[c}^{\phantom{abci}e}K_{d]e} ),
\end{equation}
that we shall use from now on in the definition of $Q$. 
Note that $Q$ uses only covariant and KY-derived objects $(f_{ab}, K_{ab},\xi^a)$, 
and the KY-generated Killing vector $\xi^a$. 

\subsection{Re-writing of $\tilde{M}_{abcd}$}

The definition \eqref{Mtilde} of the tensor $\tilde{M}_{abcd}$ can be simplified and brought into a form that is much more practical to our purposes. In particular, we can re-write the last two terms in the RHS into expressions that are fully contracted with two indices (see below). 

To begin, consider the integrability condition \eqref{intcondcentral}, and contract a free index of the Riemann tensor with the KY tensor. Expanding the anti-symmetry on the RHS gives
\begin{equation} \label{Mtildetemp1}
    f_a^{\ph e} R_{bef[c} f_{d]}^{\ph f} = \frac{1}{2} R_{cdb}^{\phantom{abm}e} K_{ae} - \frac{1}{2} R_{cdfe} f_a^{\ph f} f_b^{\ph e}.
\end{equation}
Next, anti-symmetrize over the pair $ab$ so that the LHS becomes the third term in the RHS of \eqref{Mtilde}. In the second term on the RHS of \eqref{Mtildetemp1}, insert equation \eqref{fffeps} and use the definition of the Hodge dual for the Riemann tensor to get
\begin{equation} \label{Mtildetemp2}
    f_{[a}^{\ph e} R_{b]ef[c} f_{\ph d]}^{f} = \frac{1}{2} R_{cd[a}^{\phantom{abm}e} K_{b]e} +\frac{1}{4} f_{ab} R_{cdfe} f^{fe} + \frac{\mathfrak{Z}}{2}  R_{abcd}^{\star}.
\end{equation}
Lastly, notice that the LHS of equation \eqref{Mtildetemp2} is unchanged under the exchange $ab\leftrightarrow cd$. Therefore, summing this equation with the one obtained by exchanging $ab\leftrightarrow cd$ and re-arranging gives us the following identity, valid for all Einstein spacetimes\footnote{The Einstein spacetime assumption allows us to make use of $R_{abcd}^\star = \,^\star\!R_{abcd}$, cf. Eq.~\eqref{starR-Rstar} If the spacetime is not Einstein, the formula still exist but contains extra terms.}
\begin{equation} \label{Mtildetemp3}
    f_{[a}^{\ph e} R_{b]ef[c} f_{\ph d]}^{f} =  \frac{1}{4}(R_{cd[a}^{\phantom{abm}e} K_{b]e} + R_{ab[c}^{\phantom{abm}e} K_{d]e}) +\frac{1}{8} (f_{ab} R_{cdfe} f^{fe}+f_{cd} R_{abfe} f^{fe}) + \frac{\mathfrak{Z}}{2} R_{abcd}^{\star}.
\end{equation}

Inserting this into \eqref{Mtilde} leaves us with the following simplified expression for $\tilde{M}_{abcd}$ : 
\begin{align} \label{Mtildefinal}
    \tilde{M}_{abcd} = \xi_{[a} g_{b][c} \xi_{d]} 
  + \frac{1}{2}g_{c[a} g_{b]d}\xi^2 
  -\frac{1}{16} (f_{ab} R_{cdfe} f^{fe}+f_{cd} R_{abfe} f^{fe}) - \frac{\mathfrak{Z}}{4} R_{abcd}^{\star}.
\end{align}
The point of this expression is that it admits a nicer decomposition in the bivector basis introduced in Sec. \ref{sec:biv} than its original definition \eqref{Mtilde}. The reason is that the last two terms on the RHS involve contractions of the Riemann tensor and the KY tensor over antisymmetric pairs of indices, in contrast to those in \eqref{Mtilde}.

\subsection{Conservation law in vacuum type-D spacetimes} 
\label{sec:genComDruVin}

Assuming the TD SSC, one of the goals in \cite{ComDruVin.23} was to construct a scalar quantity, denoted $\mathcal{Q}^{(2)}$ there, that is
conserved under the MPTD equations with $\kappa=1$ at \textit{quadratic-in-spin order}, in the \textit{Kerr spacetime}, using the so-called \textit{covariant building blocks} formalism (cf. Sec.~(4.1) of \cite{ComDruVin.23}).
In this section, drawing on their method, we show how their result can be extended easily to exhibit a quadratic-in-spin constant of motion for $\kappa=1$, in any \emph{Einstein spacetime endowed with a KY tensor}, using the \emph{bivector formalism} in Sec.~\ref{sec:biv}, of which the covariant building blocks formalism becomes a Kerr-adapted consequence.

A particular interesting feature of the results in \cite{ComDruVin.23}, although it only applies to Kerr, is that, up to a few (but important) steps, their methods essentially exploit geometric features that exist in all spacetimes having a KY tensor, and do not refer to Kerr-specific properties, especially if the spacetime is vacuum. Since, at least regarding the curvature tensor, Einstein spacetimes can be seen as the ``superposition'' of a vacuum type-D term (just like Kerr) and a constant-curvature term\footnote{Recall that a constant curvature spacetime is one for which the Riemann tensor is proportional to $g_{c[a}g_{b]d}$.}, we can leverage the results of \cite{ComDruVin.23} at least for the ``vacuum part'' of an Einstein spacetime. 

Table \ref{table:dictionnary} provides the dictionary between geometric quantities used in \cite{ComDruVin.23} that are specific to Kerr there, and their generalization to vacuum type-D spacetimes considered in this work. Other geometric tensors involved, namely $(g_{ab},
\varepsilon_{abcd},K_{ab},\ell^a,n^a)$ already have Kerr-independent covariant definitions in \cite{ComDruVin.23} that are the same as ours, and can be mapped to the type-D case without notation change.\footnote{The notation $N^{ab}$ to denote a bivector from \cite{ComDruVin.23} is only for this section. It is otherwise reserved for the MPTD torque \eqref{EoP} in the rest of the paper and used in Sec.~\ref{sec:quad-MPTD} only.}


\begin{table}[h]
  \centering
  \begin{tabular}{|c|c|c|c|c|}
  \hline
   & \multicolumn{2}{c}{\textbf{Reference \cite{ComDruVin.23}}} & \multicolumn{2}{|c|}{\textbf{This paper} } \\
  \hline
  \textbf{Quantity}  & Symbol & Equation & Symbol & \hspace{2mm} Equation \hspace{3mm}\\
  \hline
  Mass parameter & $M$ & -- & $ \quad- \ui \Psi_2 C^3 \quad$ & --  \\
  \hline
  Kerr scalar & $\mcR$ & (61) & $- \ui C$ & \eqref{Psi2C3cst} \\
  \hline
  KY scalar & $\mcZ$ & (53) & $\mathfrak{Z}$ & \eqref{ffstar} \\
  \hline
  NP bivector & $N_{ab}$ & (62) & $- \ui Z_{ab}$ & \eqref{defXYZ} \\
  \hline
  KY tensor & $Y_{ab}$ & (66.a) & $f_{ab}$ & \eqref{deff} \\
  \hline
  Curvature tensor & \,\, $R_{abcd}$ \,\, & (66.b) & $C_{abcd}$ & \eqref{Weyldec} \\
  \hline
  Sym. real tensor & $h_{ab}$ & (72) & $\mcZ_{ab}$ & \eqref{symbasis1} \\
  \hline
  \hspace{2mm} Bivectors identities \hspace{3mm} & -- & \hspace{2mm} (68) $\rightarrow$ (78) \hspace{3mm} & -- & Sec.~\ref{sec:biv} \\
  \hline
  \end{tabular}
  \caption{Correspondence between various geometric objects defined and used in \cite{ComDruVin.23} for the Kerr spacetime, and their generalization to vacuum, type-D spacetime used in the present work.
  }
  \label{table:dictionnary}
\end{table}


When their expressions are written in terms of the quantities on the RHS of the table above, all the formulae in \cite{ComDruVin.23} remain valid in any vacuum spacetime endowed with a KY tensor. In fact, all such formulae become corollaries of our main equations derived throughout this text. Most importantly, the proof of the central results of \cite{ComDruVin.23} relies heavily on the cornerstone identities (67) of that paper. These formulae are mapped, through the dictionary \ref{table:dictionnary}, to the following equations in our notations
\begin{subequations}\label{Fouriermap}
\begin{align} 
    \ui \nabla_c \mcR = N_{a b} \xi^b & \quad\longleftrightarrow\quad \nabla_a C = - \ui
  Z_{a b} \xi^b, \\
  \ui \nabla_c (\mcR N_{a b}) = G_{a b c d} \xi^d & \quad\longleftrightarrow\quad \nabla_c
  \Phi_{a b} = \varepsilon_{a b c d} \xi^d + 2 \ui g_{c [a}
  \xi_{b]},\\
  \nabla_a \xi_b = - \text{Im} [M \mcR^{- 2} N_{a b}] & \quad\longleftrightarrow\quad
  \nabla_a \xi_b = - \tfrac{1}{4} C_{a b c d} f^{c d}_{\star},
\end{align}
\end{subequations}
where the left-hand sides are from \cite{ComDruVin.23}, and the right-hand sides are their generalizations established in this work, in any vacuum spacetime admitting a KY tensor, by virtue of \eqref{defxi}, \eqref{dfstar} and \eqref{Dxi}. Also, note that the first line is simply the second projected onto $N_{ab}\,\,(\leftrightarrow Z_{ab})$.

With the dictionaries \ref{table:dictionnary} and \eqref{Fouriermap} in place, it is now possible to read through Sec.~(3) to Sec.~(7) of \cite{ComDruVin.23} and obtain results that are true in any vacuum type-D spacetime, and not just Kerr. This is because almost all equations used in \cite{ComDruVin.23} are analytic expressions that involve covariant definitions. The only exception is equation (103) in \cite{ComDruVin.23} which had only been checked to hold numerically using Boyer-Lindquist coordinates in Kerr. Yet, we were able to give a covariant proof for them in any vacuum type-D spacetime. Details are given in App.~\ref{app:BLtoCov}. 

\subsection{Conclusion and extension from vacuum to Einstein spacetimes}

In the above, we proved that the (generalized) Carter constant $Q$ \eqref{defQ} (with \eqref{defLabc} and \eqref{Mtildefinal}) is conserved to quadratic-in-spin order under the MPTD + TD SSC system for $\kappa=1$, in any \emph{vacuum} spacetime endowed with a KY tensor. Yet, assuming that the background is vacuum only comes from our method being based on extending the Kerr result from \cite{ComDruVin.23} to type-D spacetimes. Because Kerr is a vacuum metric, the method inherently preserved this feature.

However, our result can be made stronger than this vacuum case, as $Q$ is \emph{also} conserved for Einstein spacetimes endowed with a KY tensor. This will be proved in Sec.~\ref{sec:PBint} using a direct Poisson bracket calculation. The reason we did \emph{not} use a direct Poisson bracket calculation for the vacuum case, and relied instead on adapting \cite{ComDruVin.23}, is because i) that bracket calculation is considerably more involved owing to terms like $\nabla_a R_{bcde}$, and ii) we aimed to exploit the (almost) covariance of the covariant building blocks formalism in \cite{ComDruVin.23}, since it is precisely this that motivated our use of the bivector formalism, a natural extension of theirs.

In any case, in the Kerr spacetime and for $\kappa=1$, our invariant $Q$ defined with \eqref{defQ}-\eqref{defLabc}-\eqref{Mtildefinal} reduces exactly to the generalized Carter constant of \cite{ComDruVin.23} by construction, and covariance of all formulae involved.

\section{Covariant proof of quadratic-in-spin integrability}
\label{sec:PBint}

We are now ready to establish the main result of this paper: that the quadratic-in-spin MPTD equations of a particle endowed with the spin-induced quadrupole with $\kappa=1$ are \textit{integrable}, in any Einstein spacetime endowed with a KY tensor. This integrability follows from the Liouville-Arnold theorem \cite{Arn} and the existence of five Poisson-commuting, linearly independent constants of motion (first integrals) on the 10D Hamiltonian system constructed in Sec.~\ref{sec:Ham}. That same method was used by one of us in a Kerr background in \cite{Ra.CQG.24} with coordinate-based calculations of the Poisson brackets. Here, the result extends to Einstein spacetimes and is fully covariant. Below, Sec. \ref{subsec:Hamform} presents the definition of the five first integrals and summarizes their properties. We also explain what calculations need to be done to prove the integrability, cf. Table.~\ref{tab:poisson} Then, covariant Poisson-bracket calculations are performed in Sec.~\ref{sec:integrability}.

\subsection{Integrability} 
\label{subsec:Hamform}

\subsubsection{Statement of the integrability result} 
\label{subsubsec:result}

In any Einstein spacetime endowed with a KY tensor, the quadrupolar, quadratic-in-spin MPTD + TD SSC equations correspond to an \emph{integrable} Hamiltonian system defined on a 10-dimensional phase space when the spinning body's deformation parameter is $\kappa=1$ (black-hole-like deformation). This is ensured by the Liouville-Arnold theorem \cite{Arn,HinFle.08,Ra.Iso.IntegO1.26} and the existence of five commuting and linearly independent first integrals, denoted $(H,\Xi,\fX,K,Q)$. 

Our five first integral candidates are defined covariantly in terms of the body's momenta $(p_a,S^{ab})$, the KY tensor $f_{ab}$ and the spacetime geometry. $H$ is defined in \eqref{HquadM}. 
Associated with the Killing vectors $\xi^a=-\tfrac{1}{3}\nabla^b f^{\star a}_{b}$ and $\eta^a=K^{ab}\xi_b$ in \eqref{KSdef}, we define two Killing vector invariants 
(cf. \eqref{defFk})
\begin{equation}
    \Xi := p_a \xi^a + \frac{1}{2} S^{a b} \nabla_a \xi_b, 
    \qquad 
    \fX := p_a \eta^a + \frac{1}{2}S^{a b} \nabla_a \eta_b. 
\label{defKillInv}
\end{equation}
We also define the following two invariants following \cite{ComDruVin.23} for $K$ and Sec. \ref{sec:Carter-Rudiger} for the generalized Carter constant $Q$:
\begin{equation}
   K  := f_{a b}^{\star} S^{a b},
   \qquad 
   Q := K^{ab} p_a p_b + L_{ab}^{\phantom{ab} c} S^{ab} p_c + \tilde{M}_{abcd} S^{ab} S^{cd},   
\label{defKQ}
\end{equation}
where we recall that $K_{ab}=f_{ac}f^{c}_{\ph b}$ is the KS tensor (cf. \eqref{KSdef}), and $L_{abc}$ and $\tilde{M}_{abcd}$ were defined in \eqref{defLabc} and \eqref{Mtildefinal}, respectively. These five first integrals are functions on the 10D phase space $\mcP$, and their pairwise $\mcP$-brackets will be shown to vanish when $\kappa=1$. 

\subsubsection{About the five first integrals}

Physically speaking, the five first integrals $(H,\Xi,\fX,K,Q)$ involved in the integrability have the following properties:

\begin{itemize}\itemsep0em
    \item $H$ is the Hamiltonian, numerically equal to $-\tildemu^2/2$, with $\tildemu$ the conserved mass [cf Sec.~\ref{sec:HquadM}]. This mass is \emph{not} the dynamical mass $\mu$ (norm of the four-momentum:~\eqref{eq:def-norms}), which is not conserved at quadratic order in spin. Importantly, it is conserved for all $\kappa$, whereas $K$ and $Q$ are conserved only when $\kappa = 1$.
    \item $(\Xi,\fX)$ are two constants of motion associated with two Killing vectors $(\xi^a, \eta^a)$ constructed from the KY tensor. In Kerr, they are linked to the stationarity and axi-symmetry of the metric and are (linear combinations of) the particle's orbital energy and angular momentum. For generic Einstein type-D spacetimes, their physical interpretation depends on the metric's properties \cite{FerrSaez.04}.
    \item $K$ is a linear-in-spin constant of motion built from the KY tensor, and it is conserved both at linear order in spin \emph{and} at quadratic order in spin only when $\kappa=1$. It is often called \emph{the} R\"udiger constant as it was discovered by them in \cite{Rudiger.I.81,Rudiger.II.83}\footnote{See also \cite{ComDru.22} for a modern revisit of R\"udiger's work.}. It was shown to be conserved even at $O(2)$ with $\kappa=1$ in the Kerr spacetime in \cite{ComDruVin.23}. Its physical interpretation is the same in any metric background: the projection of the particle's spin 4-vector $S^a$ onto a covariant notion of total angular momentum \cite{DiRu.I.81,DiRu.II.82}.
    \item $Q$ is the other constant of motion built from the KY tensor. Like $K$, it was discovered first in \cite{Rudiger.I.81,Rudiger.II.83} as an attempt to generalize Carter  constant beyond Kerr geodesic cases, to linear order in spin. $Q$ was extended to quadratic order in spin in \cite{ComDruVin.23} for the Kerr spacetime, and shown to be conserved when $\kappa=1$. With these corrections, its physical interpretation is difficult, even in Kerr, and we cannot propose more than relegating it to the rank of constants coming from ``hidden symmetries'' \cite{Carter.68,Frolov.17,Yasui:2011pr}.
\end{itemize}

\subsubsection{Method for proving our result}\label{subsubsec:method}

The five constants of motion $(H,\Xi,\fX,K,Q)$ live on the reduced phase space
$\mcP$, so as in Paper~I their mutual involution must be assessed with the
Dirac bracket \eqref{Diracall} rather than the bare $\mcM$-brackets \eqref{PBs}. Beyond the
restricted bracket $\{F,G\}|_\mcP$, the Dirac bracket carries a correction
bilinear in $\{F,C^\alpha\}$ and $\{C^\beta,G\}$, so establishing integrability
requires not only the ten brackets among the constants but also the five
brackets of each constant with the constraint $C^\alpha$ feeding this
correction --- $\binom{6}{2}=15$ in total, one for each pair of distinct
elements of $(H,\Xi,\fX,K,Q,C^\alpha)$. They are collected in
Table~\ref{tab:poisson}, grouped by the way each is shown to vanish.

\begin{table}[h]
  \centering
  \renewcommand{\arraystretch}{1.35}
  \setlength{\tabcolsep}{6pt}
  \begin{tabular}{c|ccccc|c}
    $\{\,\cdot\,,\,\cdot\,\}|_\mcP$ & $H$ & $\Xi$ & $\fX$ & $K$ & $Q$ & $C^\alpha$ \\ \hline
    $H$   & $\quad 0\quad $    & \PBind{\cite{Ra.Iso.IntegO1.26}} & \PBind{\cite{Ra.Iso.IntegO1.26}} & \PBtodo{\ref{subsubsec:PBKH}}  & \PBtodo{\ref{PBQHdetails}}  & \PBdone{\eqref{PstableflowH}} \\
    $\Xi$ & \PBna  & $\quad 0\quad $    & \PBind{\cite{Ra.Iso.IntegO1.26}} & \PBind{\cite{Ra.Iso.IntegO1.26}} & \PBind{\cite{Ra.Iso.IntegO1.26}} & \PBind{\cite{Ra.Iso.IntegO1.26}} \\
    $\fX$ & \PBna  & \PBna  & $\quad 0\quad $    & \PBind{\cite{Ra.Iso.IntegO1.26}} & \PBind{\cite{Ra.Iso.IntegO1.26}} & \PBind{\cite{Ra.Iso.IntegO1.26}} \\
    $K$   & \PBna  & \PBna  & \PBna  & $0$    & \PBtodo{\ref{subsubsec:PBKQ}}  & \PBnz{~\ref{sec:shortcut}} \\
    $Q$   & \PBna  & \PBna  & \PBna  & \PBna  & $0$    & \PBtodo{\ref{sec:PBQC}} \\
  \end{tabular}
  \caption{Pairwise brackets $\{\,\cdot\,,\,\cdot\,\}|_\mcP$ on $\mcP$ among the constants $H,\Xi,\fX,K,Q$ ($5\times5$ block) and with the constraint $C^\alpha$ (last column). Since $\{F,C^\alpha\}|_\mcP=0$ for $F\in(H,\Xi,\fX,Q)$, the Dirac correction never contributes and $\{\,\}^\mcP=\{\,\}|_\mcP$ on every pair, as argued in \ref{sec:shortcut}. The table gives references for the calculation of each bracket, and the color indicates the status of the proof at this stage in the paper: \colorbox{green!12}{done directly in this paper}, \colorbox{blue!10}{inherited from Paper~I~\cite{Ra.Iso.IntegO1.26}}, \colorbox{orange!18}{done below (Sec.~\ref{sec:integrability})}, \colorbox{gray!18}{nonzero, but does not contribute (cf.~\ref{sec:shortcut})}. By antisymmetry of Poisson brackets, the lower triangle (\textendash) is redundant, and the diagonal vanishes identically.}
  \label{tab:poisson}
\end{table}

Table~\ref{tab:poisson} reduces the problem considerably. The nine blue cells --- every bracket involving one of the Killing invariants $\Xi$ or $\fX$ --- vanish under the TD SSC \eqref{TDSSC} as a consequence of the Killing-invariant identities established in Paper~I \cite{Ra.Iso.IntegO1.26}; the argument carries over from the linear-in-spin to the quadratic-in-spin case, since these brackets rely only on the isometry structure and are insensitive to the specific form of $H$.
The single green cell $\{H,C^\alpha\}|_\mcP=0$ expresses the stability of $\mcP$ under the Hamiltonian flow, established by the explicit $\mcP$-bracket calculation \eqref{PBCH}--\eqref{PBCC_PBCTheta} summarized in \eqref{PstableflowH}. The lone grey cell $\{K,C^\alpha\}|_\mcP\neq 0$ is harmless: since $K$ is the only constant with a nonzero constraint bracket and never pairs with itself, the Dirac correction to any pair $(F,G)$ is always killed by a vanishing factor from the partner constant (cf.~\ref{sec:shortcut}).

Integrability therefore hinges on the four orange cells, namely
\begin{equation} \label{eq:fourbrackets}
    \{K,H\}|_\mcP, \quad \{Q,H\}|_\mcP, \quad \{Q,K\}|_\mcP, \quad \{Q,C^\alpha\}|_\mcP,
\end{equation}
all of which we compute in Sec.~\ref{sec:integrability}. The first two are shown to
vanish only $\kappa = 1$, the calculation relying heavily on the identities
verified by the KY tensor and on the algebraic structure of type-D Einstein
spacetimes established in Sec. \ref{sec:KY}.

\subsubsection{A comment on the degenerate case $\mathfrak{Z} = 0$.}

Our integrability proof relies on the non-degeneracy assumption $\mathfrak{Z} := \tfrac{1}{4} f^\star_{ab} f^{ab} \neq 0$, which guarantees that $f_{ab}$ has rank four (we follow the nomenclature introduced in \cite{DiRu.I.81,DiRu.II.82}). It is worth commenting on what happens when this assumption fails. When $\mathfrak{Z} = 0$, the KY tensor is \emph{simple} (rank two), and by the classification of Dietz and R\"udiger~\cite{DiRu.I.81,DiRu.II.82} the spacetime inherits distinctive curvature and symmetry properties: Petrov type~D when $f_{ab}$ is non-null and type~N when null \cite{DiRu.II.82}. Two opposite situations then arise, which we address in turn: generically the symmetry content is \emph{reduced}, and integrability is lost; but in special cases it is \emph{enhanced}, and integrability survives for a different reason. The latter includes the Schwarzschild spacetime---and more generally the $a \to 0$ limit of Kerr-type geometries, where $\mathfrak{Z} \propto a$ vanishes.

In the generic degenerate case, a simple $f_{ab}$ generates no second independent Killing vector, since $\eta^a = K^{ab}\xi_b$ reduces to $\eta^a \propto \xi^a$ \cite{DiRu.I.81}. The invariant $\fX$ then loses its independence from $\Xi$, leaving only four candidates $(H,\Xi,K,Q)$: integrability is \emph{not} secured when $\mathfrak{Z}=0$. What survives is the (now reducible) KS tensor $K_{ab} = f_{ac}f^c{}_b$ and its associated invariant $Q$.

Schwarzschild shows how integrability can nonetheless persist when the symmetry is enhanced. There $\eta^a = 0$ and $\fX$ vanishes, but the geometry admits all $SO(3)$ Killing vectors~\cite{Poi}
\begin{align} \label{SO3KV}
    k_{(x)}^a &:= -\sin\phi\,(\partial_\theta)^a - \cos\phi\cot\theta\,(\partial_\phi)^a\,, \\
    k_{(y)}^a &:= \cos\phi\,(\partial_\theta)^a - \sin\phi\cot\theta\,(\partial_\phi)^a\,, \\
    k_{(z)}^a &:= (\partial_\phi)^a\,.
\end{align}
Their Dixon invariants \eqref{defFk}, denoted $(L_{x},L_{y},L_{z})$---cf. Sec.~I.C.2 of~\cite{Ra.Iso.PapII.24}, are separately conserved, so one might expect six commuting integrals: $H$, $\Xi$, the three $L_i$, and $Q$. 
%
%
However, the $L_i$ obey the $\mathfrak{so}(3)$ algebra and do not mutually commute; the maximal involutive subset is $\{L_z,\, L^2\}|_\mcP$, comprising the axial and total angular momenta, $L^2 := L_x^2 + L_y^2 + L_z^2$.
Second, the Carter-type constant $Q$ is not independent: one finds $Q = -L^2$ in Schwarzschild (cf. the companion notebook \cite{MMA_IntegO2}). The commuting set is therefore $(H, \Xi, L_z, L^2, K)$: $L_z$ replaces the lost $\fX$ and $L^2$ reproduces $Q$, restoring the fifth integral. This rescue owes to the enhanced $SO(3)$ symmetry, not to $\mathfrak{Z}=0$ itself. As in Kerr (and elsewhere in this work), quadratic-in-spin integrability in Schwarzschild holds for a black-hole secondary ($\kappa = 1$), for which the R\"udiger constant $K$ is conserved.

\subsection{Computation of the remaining Poisson brackets} 
\label{sec:integrability}

For the constants of motion $(\Xi,\fX)$ in \eqref{defKillInv} 
(associated with Killing vectors $(\xi^a, \eta^a)$), the same argument used in the linear-in-spin case applies as in Paper~I: with quasi-symplectic coordinates, the invariant commutes with any function of phase space coordinate built from the metric and momenta $(p_a, S^{ab})$. Evidently, it does not commute with its conjugated coordinate, but by definition, this coordinate does not appear in the metric coefficients. 
We thus have to show that the remaining two constants of motion $Q,K$ (built from the KY tensor) commute with $H$, under the $\mcP$-brackets \eqref{PBs}.

\subsubsection{Proof of $\{K, H\}|_\mcP=0$ for $\kappa=1$} \label{subsubsec:PBKH}

We start with the first integral $K$ defined in \eqref{defKQ}. Using the Hamiltonian \eqref{HquadM}, the brackets \eqref{PBs} and the decomposition \eqref{zip!}, we find
\begin{equation} \label{PBKH}
    \{K,H\} = \nabla_\alpha f_{\beta\gamma}^{\star} p^\alpha S^{\beta\gamma} + 4 R_{\alpha\beta[\gamma}^{\ph\ph\ph\,\, \lambda} f_{\delta]\lambda}^{\star} \hat{p}^\alpha S^\beta \hat{p}^\gamma S^\delta + (1-\kappa) R_{\alpha\beta\gamma}^{\ph\ph\ph \lambda} f_{\lambda\delta}^{\star} \hat{p}^\alpha S^\beta \hat{p}^\gamma S^\delta+ O (3,C^\alpha),
\end{equation}
where we also used the intermediate brackets 
$\{ S^{\lambda \mu}, \Theta^{\beta \gamma} \} = 2 (g^{\lambda(\gamma} \Theta^{\beta)\mu} - \Theta^{\lambda (\gamma} g^{\beta)\mu})$ and $\{ S^{\lambda\mu}, C^\delta \} =2 p^{[\lambda} S^{\mu] \delta} + O (2,C^\alpha)$. The first term in \eqref{PBKH} is the linear-in-spin contribution which vanishes for any spacetime with a KY tensor (as shown in \cite{Ra.Iso.IntegO1.26}). The second term vanishes in virtue of its algebraic symmetries and the KY integrability condition \eqref{intcondcentral_star}. The third term, however, does not vanish in general unless $\kappa=1$. This is already true in Kerr, as shown in \cite{ComDruVin.23}. As a consequence, we have shown that $\{K,H\}|_\mcP=0$ for $\kappa=1$. 

We also note that in the case of a degenerate KY tensor, i.e., when $\frak{Z}=0$, as in the Schwarzschild spacetime, the conclusions on the RHS of equation \eqref{PBKH} still apply. In particular, the first and second term vanish (this is solely based on \eqref{dfstar} and \eqref{intcondcentral_star} which hold for any KY tensor, not just non-degenerate ones), and the third term vanishes for $\kappa=1$, but not for $\kappa\neq 1$. 

\subsubsection{Proof of $\{Q,H\}|_\mcP=0$ for $\kappa=1$} \label{PBQHdetails}

Next, the first integral $Q$ defined in \eqref{defKQ}. Using the Hamiltonian \eqref{HquadM} with $\kappa=1$ and the brackets \eqref{PBs}, a long but otherwise straightforward calculation gives
\begin{equation} \label{HQtot}
  \{ Q,H \} = \left( 
    g_{\alpha\delta} K_\nu^{\ph \rho} \nabla_\rho R_{\lambda\beta\gamma\mu} 
  + \tfrac{1}{2} g_{\lambda\mu} L_{\alpha\beta}^{\ph\ph \rho} R_{\gamma\delta\nu\rho} 
  + g_{\lambda \mu} \nabla_\nu \tilde{M}_{\alpha \beta\gamma\delta} 
  \right) 
  S^{\alpha \beta} S^{\gamma \delta} \hat{p}^\lambda \hat{p}^\mu p^\nu,
\end{equation}
where we have already removed the linear-in-spin contributions, which was shown to vanish in Paper I.
Equation \eqref{HQtot} holds for any spacetime admitting a KY tensor. We restrict to Einstein spacetimes by inserting the decomposition \eqref{fullcurv}. Since the RHS of \eqref{HQtot} is linear in $R_{abcd}$ (including terms contained in $\tilde{M}_{abcd}$), this substitution will give two independent contributions:
\begin{itemize}
    \item one piece identical to \eqref{HQtot} with $R_{abcd}\rightarrow C_{abcd}$. By construction, this corresponds to $\dot{Q}$ in a vacuum spacetime (recall \eqref{HamEqM}) and vanishes following our work in Sec.~\ref{sec:Carter-Rudiger}.  
    \item one piece proportional to $\Lambda$ and amounts to setting $R_{abcd}\rightarrow\tfrac{2}{3} \Lambda g_{a[c}g_{b]d}$ in \eqref{HQtot}. This contribution vanishes too, as shown in the following paragraphs.
\end{itemize}

When substituting $R_{abcd}\rightarrow\tfrac{2}{3}\Lambda g_{a[c}g_{b]d}$ into \eqref{HQtot}, the first and second terms are readily seen to be $O(C^a)$ and vanish identically, respectively. Consequently, we are left with a single remaining contribution to the RHS of \eqref{HQtot}, namely the $O(\Lambda)$ contribution from the first term. Using equations \eqref{fullcurv}, \eqref{Mtildefinal} and \eqref{Dxi}, equation \eqref{HQtot} becomes
\begin{equation} \label{HQred}
    \{Q,H\} = - \frac{\Lambda}{3}\left( (2 f^{\star}_{\nu\alpha} \xi_\beta - g_{\alpha \beta} f^{\star}_{\nu\lambda} \xi^\lambda)\Theta^{\alpha\beta} - \tfrac{1}{4} (2 f_{\alpha\beta} \nabla_\nu f_{\gamma \delta} + \nabla_\nu \mathfrak{Z} \varepsilon_{\alpha\beta\gamma\delta})S^{\alpha\beta} S^{\gamma\delta} \right)  p^\nu\,,
\end{equation}
where in the first line we used $g_{ef}\hat{p}^e\hat{p}^f=-1$, the symmetries of $S^{ab}S^{cd}$ and the fact that $\Lambda, g_{ab}$ and $\varepsilon_{abcd}$ are covariantly constant. Using equations \eqref{SsunderSSC}, \eqref{defxi}, and applying \eqref{TDSSC}, we find that equation \eqref{HQred} reduces to 
\begin{equation} \label{QH}
    \{Q,H\}= -\frac{2 \Lambda}{3} \left(S_{\circ}^2 \hat{p}^\alpha \hat{p}^\beta f^{\star}_{\alpha \nu} \xi_\beta + S^\alpha S^\beta f^{\star}_{\nu\alpha} \xi_\beta + 2 \hat{p}_{[\nu} S_{\rho]}\xi^\rho f^{\star}_{\gamma \delta} \hat{p}^\gamma S^\delta \right) p^\nu + O(C^\alpha).
\end{equation}
In the RHS of this equation, expanding the antisymmetry in $\hat{p}_{[g}S_{h]}$ gives four terms. The second and third cancel each other, the first one vanishes by anti-symmetry of $f^\star_{ag}$, and the fourth is $\propto S_g p^g$ which vanishes by construction (cf. \eqref{zip!}). Consequently, $\{Q,H\}=O(C^\alpha)$ and we can conclude, as claimed, that $\{Q,H\}|_\mcP=0$ (for $\kappa = 1$).

\subsubsection{Proof of $\{K,Q\}|_\mcP=0$} \label{subsubsec:PBKQ}

Using the definitions \eqref{defKQ} and the brackets \eqref{PBs}, the bracket $\{K,Q\}$ is given by
\begin{align}
    \{ K,Q \} &= 2 \left( K_\gamma^{\ph \delta} \nabla_\delta f^{\star}_{\alpha\beta} + 2 f^{\star \delta}_\beta L_{\delta\alpha\gamma} \right) S^{\alpha\beta} p^\gamma \label{PBKQ} \\
    &\quad+ ( L_{\gamma \delta}^{\phantom{\gamma \delta} \lambda} \nabla_\lambda f_{\alpha \beta}^{\star} - 8f_\alpha^{\star \lambda} \tilde{M}_{\lambda \beta \gamma \delta}) S^{\alpha \beta} S^{\gamma \delta} \notag + O(C^\alpha).
\end{align}

Again, the linear-in-spin term in the first line has been shown to vanish in Paper I. For the quadratic-in-spin terms in the second line, we first consider the two contributions from $L_{a b c}$ and $\tilde{M}_{a b c d}$ separately. Using the definitions \eqref{defLabc}--\eqref{Mtildefinal}, the decomposition \eqref{zip!}, \eqref{TDSSC} and various KY identities from Sec. \ref{sec:KY}, we obtain
\begin{subequations} \label{LSSMSS}
    \begin{align}
    L_{cd}^{\phantom{cd} e} \nabla_e f^{\star}_{ab} S^{a b} S^{c d} &= 4f^{\star}_{c e} \xi^e \xi_b \Theta^{b c}, \\
    8 f_a^{\star e} \tilde{M}_{ebcd} S^{a b} S^{cd} & = 4 f^{\star}_{c e} \xi^e \xi_b \Theta^{b c} - 4 f^{\star}_{c a} S^{a b} \xi_b S^{c d} \xi_d \label{2ndlinefMSS} \\
    &- \tfrac{1}{2} (f^{\star e}_c f_{e d} R_{a b f g} f^{f g} + f_{a b} f^{\star e}_c R_{e d f g} f^{f g} + 4 \mathfrak{Z} f^{\star e}_c R_{e d a b}^{\star}) S^{a b} S^{c d}. \notag
    \end{align}
\end{subequations}
with the second term in \eqref{2ndlinefMSS} vanishing, since $f_{ca}=f_{[ca]}$ is contracted twice with the same vector $S^{ab} \xi_b$. Upon inserting \eqref{LSSMSS} into \eqref{PBKQ}, the terms involving $\xi^e$ cancel each other. The remaining terms involving the Riemann tensor and its dual are then split into a Weyl and a $O (\Lambda)$ pieces using equation \eqref{fullcurv}, as we did for $\{Q,H\}$. We then have the two contributions as
follows:
\begin{align} \label{KQbis}
    \{ K, Q \} &= - (f^{\star \lambda}_\gamma f_{\lambda \delta} C_{\alpha \beta \mu \nu} f^{\mu \nu} + f_{\alpha \beta} f^{\star \lambda}_\gamma C_{\lambda \delta \mu \nu} f^{\mu \nu} + 4 \mathfrak{Z} f^{\star \lambda}_\gamma C_{\lambda \delta \alpha \beta}^{\star}) S^{\alpha \beta} S^{\gamma \delta}\\
    & \qquad - \frac{\Lambda}{3} (f^{\star \lambda}_\gamma f_{\lambda \delta} f_{\alpha \beta} + f_{\alpha \beta} f^{\star \lambda}_\gamma f_{\lambda \delta} + 2 \mathfrak{Z} f^{\star \lambda}_\gamma \varepsilon_{\alpha \beta \lambda \delta}) S^{\alpha \beta} S^{\gamma \delta} + O (C^\alpha). \notag
\end{align}
These two lines can be shown to vanish independently by using the following identities, obtained from a combination of equations \eqref{ffstar}, \eqref{fffeps}, \eqref{fstarNP} and \eqref{CXYZ}:
\begin{subequations}
    \begin{align}
        f^{\star e}_c C_{e d a b}^{\star} & = \text{Re} [\Psi C (X_{a b} Y_{c d} - X_{c d} Y_{a b})],\\
        f^{\star e}_c \varepsilon_{abe d} & = 2 g_{c [a} f_{b] d}- f_{ab} g_{c d}, \\
        f^{\star e}_c C_{e d f g} f^{f g} & = 2 \text{Im} [\Psi C^2] g_{c d}.
    \end{align}
\end{subequations}
Inserting these identities and \eqref{ffstar} into \eqref{KQbis} readily implies that both lines vanish, as all terms boil down to combinations $g_{c d} S^{c d}$, $(X_{a b} Y_{c d} - X_{c d} Y_{a b}) S^{a b} S^{c d}$ and $f_{a b} \Theta^{a b}$, 
all of which vanish: the first and third by contraction of symmetric and antisymmetric index pairs, and the second by antisymmetry under $ab \leftrightarrow cd$ against the symmetric $S^{a b} S^{c d}$. Consequently, we have shown as claimed that $\{ K, Q \} |_\mcP = 0$.

\subsubsection{Proof of $\{Q,C^\alpha\}|_\mcP=0$} \label{sec:PBQC}

Repeated uses of definitions \eqref{defKQ}, brackets \eqref{PBs} and the Leibniz rule lead to the following expression
\begin{align}\label{PBQC}
    \{ Q, C^\lambda \} &= (\nabla_\gamma K_{\alpha \beta} - 2 L_{\gamma \alpha \beta}) \,p^\alpha p^\beta S^{\gamma \lambda} \\ 
    &\quad+ ( R_{\alpha \beta \delta \mu} K_{\phantom{\beta} \gamma}^\mu + \nabla_\delta L_{\alpha \beta \gamma} + 4\tilde{M}_{\alpha \beta \gamma \delta} ) \,S^{\alpha \beta} p^\gamma S^{\delta \lambda} + O (C^\alpha). \notag
\end{align}
    
The linear-in-spin term was shown to be $O(C^\alpha)$ in any spacetime admitting a KY tensor in Paper I \cite{Ra.Iso.IntegO1.26}. We show that the quadratic-in-spin piece in the second line vanishes in Einstein type-D spacetimes as follows. Let us define the following tensor:
\begin{equation} \label{Fabcd}
    F_{abcd}:=R_{abde}K^e_{\ph c}+\nabla_d L_{abc}=R_{abde}K^e_{\ph c} + \frac{2}{3}\nabla_d \nabla_{[a}K_{b]c} +\frac{4}{3} \varepsilon_{a b c f} \nabla_d \nabla^f \mathfrak{Z},
\end{equation}
already introduced in Paper I. There, it was shown that $F_{abcd}$ has the same symmetries as $\tilde{M}_{abcd}$, namely \eqref{symM}. Using this definition, the bracket \eqref{PBQC} reads
\begin{equation} \label{QCalt}
    \{Q,C^\lambda\}= ( F_{\alpha\beta\gamma\delta} + 4\tilde{M}_{\alpha \beta \gamma \delta} ) \,S^{\alpha \beta} p^\gamma S^{\delta \lambda} + O (C^\alpha,3).
\end{equation}
Notice that the symmetries of $F_{abcd}$ and $\tilde{M}_{abcd}$ do not imply that either of these tensors is anti-symmetric in $bc$. However, quite remarkably, \emph{their sum} does possess that symmetry:
\begin{equation} \label{Fa(bc)d=0}
    F_{a(bc)d} + 4\tilde{M}_{a(bc)d}=0. 
\end{equation}
This property is largely non-trivial, and was found serendipitously using coordinate calculations in Kerr. The proof is detailed in App.~\ref{app:Fa(bc)d} and exploits the structure of the Riemann tensor (\eqref{fullcurv}--\eqref{CXYZ}), and decompositions within the bivector formalism. 

Now we make the following point: since the tensor $F_{abcd}+4\tilde{M}_{abcd}$ has the combined symmetries \eqref{symM} and \eqref{Fa(bc)d=0}, it must be totally anti-symmetric, i.e. a 4-form (the proof is just a matter of manipulating indices). Consequently, since 4-forms in 4-dimensional spacetimes span a 1-dimensional vector space, $F_{abcd}+4\tilde{M}_{abcd}$ \emph{must} be proportional to the Levi-Civita tensor $\varepsilon_{abcd}$ (another 4-form). Using Eqs.~\eqref{epsIds}, the (algebraic) Bianchi identity, allows one to find this explicit relation, which reads
\begin{equation} \label{FMeps}
    F_{abcd} + 4\tilde{M}_{abcd} = \frac{1}{2} \,\Delta\mathfrak{Z} \,\varepsilon_{abcd}, 
\end{equation}
where $\Delta=g^{ab}\nabla_a\nabla_b$. In Einstein spacetimes, the scalar field $\Delta\mathfrak{Z}$, though constructed solely from the KY tensor, contains information about both the Weyl part and the $O(\Lambda)$ part of the curvature. This can be seen by taking two covariant derivatives of \eqref{ffstar} and using \eqref{defxi}, \eqref{dfstar} and \eqref{Dxi}, leading to
\begin{equation} \label{DZ}
    \frac{1}{2}\Delta \mathfrak{Z} 
    = 
    \frac{1}{8} C_{abcd} f^{a b} f^{c d}_{\star} - \frac{2 \Lambda}{3} \mathfrak{Z}.
\end{equation}
By inserting \eqref{FMeps} into \eqref{QCalt} and the decomposition \eqref{zip!}, we finally obtain
\begin{equation} \label{QCaltfinal}
    \{Q,C^\lambda\} = \frac{1}{2} \,\Delta\mathfrak{Z} \,\varepsilon_{\alpha\beta\gamma\delta}\,S^{\alpha \beta} p^\gamma S^{\delta \lambda} + O(C^\alpha).
\end{equation}
Under the TD SSC, substituting \eqref{TDSSC} into the first term and applying the contraction identity \eqref{epsIds} yields a term proportional to $S_\delta S^{\delta\lambda}$; substituting \eqref{TDSSC} once more gives $S_\delta S_\sigma \varepsilon^{\delta\lambda\rho\sigma}\hat{p}_\rho$, which vanishes by symmetry of $S_\delta S_\sigma$ against the antisymmetry of $\varepsilon^{\delta\lambda\rho\sigma}$.
Hence $\{Q,C^\alpha\}|_\mcP = 0$, and this concludes our proof of the integrability result \ref{subsubsec:result}. 

\begin{acknowledgments}
This work benefited from constant interactions with many colleagues at all stages. 
We thank A.~N.~Seenivasan, S.~Dolan, J.~Mathews, A. Pound,  M.~Rahman, M.~Shahzadi, and V.~Witzany for useful comments and feedback. We thank G. Comp\`{e}re for constructive feedback on Sec~\ref{sec:KY}, and for clarifying the relevant references \cite{ComDru.22,ComDruVin.23}. 
P.R. is indebted to A.~Harte and L.~Stein for their assistance with the Newman-Penrose and bivector formalism. P.R. also acknowledges the hospitality of Universit\'e Libre de Bruxelles where part of this work was conducted.
S.I. thanks T.~Houri for enlightening discussions on the classification of the type-D spacetimes and guidance through the relevant literature, and A.~J.~K. Chua for his continuous encouragement. 
Finally, we gratefully acknowledge the Institute for Mathematical Sciences at the National University of Singapore for hosting the workshop `Mathematical Methods for the General Relativistic Two-body Problem' (August 2025), during which the final part of this work was completed, and for their warm hospitality.
P.R. acknowledges funding from the France 2030 program, as part of the 3NC project. S.I. is supported by the Ministry of Education, Singapore, under the Academic Research Fund Tier 1 A-8001492-00-00 (FY2023).
\end{acknowledgments}

\bibliography{ListeRef, ListeRef_Sis}

\appendix

\section{Conventions, notations} \label{app:ConvNot}

Throughout the paper, we use geometric units in which $G=c=1$ with the metric signature $(-,+,+,+)$. Abstract indices are $a,b,c,\dots$; coordinate components use Greek indices $\alpha,\beta,\dots$; and orthonormal tetrad indices are $A,B,\dots$. We denote the metric tensor, the Levi-Civita (totally anti-symmetric) tensor and the covariant derivative by $g_{ab}$, $\varepsilon_{abcd}$ and $\nabla_a$, respectively.
Symmetrization and antisymmetrization are
$T_{(ab)} := \tfrac12\,(T_{ab} + T_{ba})$
and 
$T_{[ab]} := \tfrac12\,(T_{ab} - T_{ba})$, respectively; 
vertical bars exclude indices, e.g.
$T_{(a|b|c)}=\tfrac12\,(T_{abc}+T_{cba})$ 
and 
$T_{[ab|c|d]}=\tfrac12\,(T_{abcd}-T_{bacd})$.

Our curvature conventions follow Wald \cite{Wald}:
$\nabla_{a}\nabla_b \omega_c = R_{abc}^{\phantom{abc}d}\omega_d$ 
for any 1-form $\omega_a$ 
and 
$
G_{ab} := R_{ab} - (1/2) R\,g_{ab}.
$
Einstein ($\Lambda$‑vacuum) spacetimes satisfy $T_{ab}=0$, equivalently $R_{ab} = \Lambda g_{ab}$; Ricci‑flat (vacuum) spacetimes have $\Lambda=0$ and $R_{ab}=0$.

On phase space we write $\{ \cdot , \cdot \}$ for the Poisson bracket on the 14D phase space $\mcM$, and $\{ \cdot , \cdot \}^{\mcP}$ for the Dirac bracket on the 10D physical space $\mcP$. We use RHS and LHS as shorthands for right-hand side and left-hand side, respectively. A table for other frequently used symbols, description and references is given in Table.~\ref{Table}.

\begin{table}[!ht]
    \caption{List of frequently used symbols.}
    \vspace{0.2cm}
	\begin{tabular}{lll}
		\toprule
        \textbf{Notation}         & \textbf{Description}                  & \textbf{Reference}  \\
        \midrule  
		\textbf{Spacetime}      &                                     &     \\
		$R_{abcd}, C_{abcd}$    & Riemann/Weyl curvature tensor               &     \eqref{fullcurv}, \eqref{CXYZ}      \\
        $\Psi_2$              & Weyl curvature scalar (type-D)              &     \eqref{defPsi2}, \eqref{eq:C_Psi}       \\
        $R,\Lambda$             & Ricci scalar / Cosmological constant                  &    \eqref{defLambdaphi2}     \\
        $(l^a,n^a,m^a,\bar{m}^a)$              & complex null tetrad               &      \eqref{orthonlm}    \\
        $(X_{ab},Y_{ab},Z_{ab},\ldots)$              & complex bivectors               &     \eqref{defXYZ}     \\
        $(\mcX_{ab},\mcY_{ab},\mcZ_{ab},\ldots)$     & complex symmetric tensors               &      \eqref{symbasis1}, \eqref{symbasis2}    \\
        \midrule     
  
		\textbf{Symmetries}         &                                      &         \\
        $f_{ab}$                & Killing-Yano tensor                   &  \eqref{deff}, \eqref{fNP} \\
        $\Phi_{ab}$                & complex conformal Killing-Yano tensor                    &  \eqref{defPhi} \\
        $(\mathfrak{Z},C)$                & Killing-Yano scalars                    & \eqref{deff},  \eqref{KYnorms}  \\
        $K^{ab}$                & Killing-St\"ackel tensor                &  \eqref{KSdef}, \eqref{decompKS}        \\
        $(\xi^a,\eta^a)$          & (KY-based) Killing vectors                       & \eqref{KSdef}         \\
        $(\Xi,\mathfrak{X})$                   & Killing vector invariants                  &  \eqref{defKillInv}         \\
		$Q$               & (generalised) Carter constant                       &  \eqref{defQ}        \\
		$K$               & R\"udiger constant                   &  \eqref{defKQ}         \\
		\midrule     
  
		\textbf{Particle}         &                                      &         \\
        $p_a$               & momentum 1-form              &     \eqref{IspSJ}, \eqref{EoM}     \\
		$S^{ab}$                & antisymmetric spin tensor                &    \eqref{IspSJ}, \eqref{EoP}   \\
        $J^{abcd}$              & spin-induced quadrupole           &       \eqref{IspSJ}, \eqref{EEgen}, \eqref{Jabcd}     \\
        $\kappa$                & spin-induced deformability parameter      &   \eqref{Jabcd}  \\ 
        $F^a,N_{ab}$            & quadrupolar force and torque &         \eqref{FNquad}      \\
		$\mu,S_\circ,S_\star$         & dynamical mass, spin norms       &  \eqref{eq:def-norms}, \eqref{SsunderSSC}        \\
        $u^a$           & particle's four-velocity       &  \eqref{momvelTD}      \\
        $\tilde{\mu}$                     & conserved (effective) mass        &   \eqref{mutilde} \\
        $S^a,C^a$     & spin and mass dipole wrt $p^a$     &       \eqref{zip!}    \\
        TD SSC     & Tuclzyjew-Dixon spin supplementary condition     &       \eqref{TDSSC}    \\
        
        \midrule
        
		\textbf{Phase space}           &                                &          \\
		$\mcM$                    & 14D phase-space manifold            &  \ref{subsec:gens}         \\
 		$(x^\alpha,p_\alpha,S^{\alpha\beta})$                    & coordinates on $\mcM$          &  \eqref{PScoord}         \\
        $\mcP$                  & 10D physical phase space               &  \eqref{defP}         \\
        $H$                     & Hamiltonian on $\mcM$               &  \eqref{HquadM}      \\
        $\lambda_H $     & Hamiltonian ``time'' parameter              &  \eqref{lambdaH}      \\
        $\{\,,\}$                 & Poisson brackets on $\mcM$                   &  \eqref{PBs}        \\
        $\{\,,\}^\mcP$         & Poisson brackets on $\mcP$                    &  \eqref{Diracall}         \\
        $\{\,,\}|_\mcP$                 & Poisson brackets on $\mcM$ evaluated on $\mcP$                    &      \eqref{KeyResult}     \\
        \bottomrule
	\end{tabular}
    \label{Table}
\end{table}

\section{Kerr-de Sitter values} 
\label{App:KdS}

The results presented in this work apply to Einstein spacetimes that admit a Killing-Yano tensor. The canonical example of this class is the Kerr-de Sitter spacetime, characterized by mass $M$, spin parameter $a$, and cosmological constant $\Lambda$.

This appendix lists the components of key geometric tensors used in our analysis, expressed in the basis associated with Boyer-Lindquist coordinates $(t,r,\theta,\phi)$. For further details, including additional expressions and verification of the coordinate calculations, we refer the reader to the accompanying Mathematica notebook~\cite{MMA_IntegO2}.

\subsection{Symbols and metric}

We start with the following useful symbols used throughout this appendix and the Mathematica companion notebook: 
\begin{subequations}
    \begin{align}
    \varpi&:= a^2+r^2, \quad
    \Sigma := r^2+a^2\cos^2\theta, \quad 
    \rho := r + \ui a \cos \theta, \\
    \lambda &:= \Lambda/3, \quad 
    \chi := 1 + \lambda a^2 , \quad
    \sigma := \sin\!\theta, \\
    \Delta_r & := \varpi \left(1-\lambda  r^2\right)-2 M r ,\quad
    \Delta_\theta := 1+ \lambda a^2 \cos^2\theta,
    \end{align}
\end{subequations}
all of which are real-valued except for $\rho\in\CC$. With these notations, the metric tensor $g_{ab}$ of the Kerr-de Sitter spacetime has components $g_{\alpha\beta}$ given (in matrix notation) by
\begin{equation} \label{comp_metric}
    g_{\alpha\beta}= \frac{1}{\Sigma\chi^2} \left(
\begin{array}{cccc}
 a^2 \sigma^2 \Delta_{\theta} -\Delta_r & 0 & 0 & -a \sigma^2 \left(\varpi \Delta_{\theta}-\Delta_r\right) \\
 0 & \Sigma^2\chi^2/\Delta_r & 0 & 0 \\
 0 & 0 & \Sigma^2\chi^2/\Delta_{\theta} & 0 \\
 -a \sigma^2 \left(\varpi \Delta_{\theta}-\Delta_r\right) & 0 & 0 & \sigma^2 \left(\varpi^2 \Delta_{\theta}-a^2 \sigma^2 \Delta_r\right) \\
\end{array}
\right).
\end{equation}

\subsection{Null tetrad and bivectors}

A tetrad of null vectors can be constructed in the Kerr-de Sitter spacetime following classical methods \cite{Ki.69}. The following components make up the complex-valued vectors $(l^a,n^a,m^a,\bar{m}^a)$ :
\begin{subequations}
\begin{align}
  l^{\alpha} & = \frac{1}{
  \Delta_r} \, \left( \chi \varpi, \Delta_r, 0, a \chi \right), \\
  n^{\alpha} & = \frac{1}{2 \Sigma} \, (\chi \varpi, - \Delta_r, 0, a \chi), \\
  m^{\alpha} & = \frac{1}{\sqrt{2 \Delta_{\theta}} \, \bar{\rho}} \, (- \ui  a \chi \sigma, 0, \Delta_{\theta}, - \ui 
  \chi/\sigma), \\
  \bar{m}^{\alpha} & = \frac{1}{\sqrt{2 \Delta_{\theta}} \, \rho} \, (\ui  a \chi \sigma, 0, \Delta_{\theta}, \ui 
  \chi/\sigma).
\end{align}
\end{subequations}

From this null tetrad, one constructs the bivectors as presented in equation \eqref{defXYZ}. The bivector $Z_{ab}$ is particularly relevant for our purposes. Its complex-valued components are given by 
\begin{equation} \label{comp_Z}
  Z_{\alpha \beta} = \chi^{- 1} \left( 
  \begin{array}{cccc}
    0 & 1 & \ui a \sigma & 0\\
    - 1 & 0 & 0 & a \sigma^2\\
    - \ui a \sigma & 0 & 0 & \ui \varpi \sigma\\
    0 & - a \sigma^2 & - \ui \varpi \sigma & 0
  \end{array} \right).
\end{equation}

Similarly, the real-valued symmetric tensor $\mcZ_{ab}$ can be constructed according to \eqref{symbasis1}, and its components are given by 
\begin{equation} \label{comp_mcZ}
  \mcZ_{\alpha \beta} = \frac{1}{\Sigma\chi^2} \left( \begin{array}{cccc}
    - (a^2 \sigma^2\Delta_{\theta}  + \Delta_r) & 0 &
    0 & a (\varpi \Delta_{\theta} + \Delta_r)\sigma^2 \\
    0 & \Sigma^2\chi^2/\Delta_r & 0 & 0\\
    0 & 0 & - \Sigma^2\chi^2/\Delta_{\theta} & 0\\
    a (\varpi \Delta_{\theta} +
    \Delta_r)\sigma^2  & 0 & 0 & - (\varpi^2
    \Delta_{\theta} + a^2 \sigma^2 \Delta_r)\sigma^2
  \end{array} \right).
\end{equation}

\subsection{Curvature and Killing objects}

We complete the list of explicit coordinate expressions by giving the curvature scalar of the Kerr-de Sitter spacetime:
\begin{equation}
  \Psi_2 = - \frac{M}{(r - \ui a \cos \theta)^3} \quand R = 4 \Lambda;
\end{equation}
and the various Killing quantities involved in our analysis. First, the scalars 
$\mathfrak{Z}$ and $C$ defined in \eqref{deff} and {\eqref{fNP}}, respectively, are given by
\begin{equation} \label{comp_C}
    \mathfrak{Z} = a r \cos \theta
    \quand
    C = a \cos\theta +\ui r.
\end{equation}
Second, the two Killing vectors $(\xi^a,\eta^a)$ have components $\xi^{\alpha} = \chi (1, 0, 0, 0)$ and $\eta^{\alpha} = \chi (a^2, 0, 0, a)$, such that, in a Kerr-de Sitter spacetime, 
\begin{equation}
    \xi^a = (\partial_t)^a \quand \eta^{\alpha} = \chi a \bigl( (\partial_\phi)^a + a (\partial_t)^a \bigr).
\end{equation}
Lastly, the components of the Killing-Yano tensor $f_{ab}$ and the Killing-St\"{a}ckel tensor $K_{ab}$ can be obtained by combining equations \eqref{comp_metric}, \eqref{comp_Z}, \eqref{comp_mcZ}, \eqref{comp_C} and the following identities
\begin{equation}
    f_{\alpha\beta} = \text{Re} [CZ_{\alpha\beta}] \quand K_{\alpha\beta} = \frac{1}{2} \text{Re} [C^2] g_{\alpha\beta} + \frac{1}{2} | C |^2 \mcZ_{\alpha\beta},
\end{equation}
which had already been derived in Sec. \ref{sec:KY}. 

\section{Hodge duality and Levi-Civita tensor identities}
\label{app:HOdge-LevC}

\subsection{Hodge duality for 2-forms and the Riemann tensor}
\label{app:Hodge}

Consider an anti-symmetric tensor $A_{ab}$. We define its Hodge dual by
\begin{equation} \label{defHodge}
    A_{ab}^\star := \frac{1}{2} \varepsilon_{ab}^{\ph\ph cd}A_{cd}.
\end{equation}
Applying the Hodge dual one more time to \eqref{defHodge} and using identity \eqref{epsIds} below, we obtain
\begin{equation}
    A_{ab}^{\star\star} = - A_{ab},
\end{equation}
i.e., applying the Hodge dual twice is equivalent to flipping the sign in front of a 2-form.

For the Riemann tensor, the algebraic symmetries imply two distinct 2-indices contractions with the Levi-Civita tensor. They result in the so-called right- and left-Hodge duals of the Riemann tensor. They are, respectively,
\begin{equation} \label{defHstar}
    R^{\star}_{abcd} =\frac{1}{2}\varepsilon_{cdef} R_{ab}^{\ph\ph ef} \quand ^\star \!R_{abcd} =\frac{1}{2}\varepsilon_{abef}R^{ef}_{\ph\ph cd} = R^\star_{cdab},
\end{equation}
where the second equality on the RHS follows from the exchange symmetry of both $R_{abcd}$ and $\varepsilon_{abcd}$. We note that, by definition, these Hodge duals are trace-free. For example, 
\begin{equation}
  g^{ab} R^{\star}_{acbd} = - \frac{1}{2} R_{cabe} \varepsilon_{\phantom{abe} d}^{abe} = 0,
\end{equation}
where the second equality follows from the total antisymmetry of $\varepsilon^{dabe}$ and Bianchi identity $R_{c[abe]}=0$. The same result holds for the left-Hodge dual.

It is clear from \eqref{defHstar} that applying the right- and left-Hodge duals on the Riemann tensor are distinct operations in general. In fact, their difference satisfies the following identity\footnote{This is a consequence of Eq.(A.15) of \cite{ComDruVin.23}, and more generally, of a broader class of results called \textit{dimension-dependent identities}, cf. the method described in Sec. 5.1.3 of \cite{Nutma.2013}.}:
\begin{equation} \label{starR-Rstar}
    ^\star \!R_{abcd}-R^\star_{abcd}=\frac{1}{2}\varepsilon_{abcd}R+2\varepsilon_{cde[a}R^e_{\ph b]}.
\end{equation}
The RHS of ~\eqref{starR-Rstar} vanishes for vacuum spacetimes, which means that the left and right Hodge duals of the Weyl tensor are equivalent. The RHS of \eqref{starR-Rstar} also vanishes for Einstein spacetimes for which $R_{ab}=\Lambda g_{ab}$ for some constant $\Lambda$. Thus, for Einstein spacetimes, both Hodge duals coincide\footnote{In an Einstein spacetime, the scalar field $\Lambda$ defined such that $R_{ab}=\Lambda g_{ab}$ is necessarily covariantly constant by the algebraic Bianchi identity. Note also, for Einstein spacetimes, that $G_{ab} + \Lambda g_{ab} = 0$ and that the Ricci scalar is covariantly constant too since $R=4\Lambda$.}.

However, for spacetimes admitting a KY tensor, the Ricci tensor has two independent components encoded in $\Lambda$ and $\phi_2$, cf. equation \eqref{finalCandCRicci}. Therefore, for spacetimes admitting a KY tensor, equation \eqref{starR-Rstar} becomes
\begin{equation} \label{duals}
    ^\star \!R_{abcd}-R^\star_{abcd}=4\phi_2\varepsilon_{cde[a}\mcZ^e_{\ph b]}= 4 \phi_2 \text{Im} [Z_{a b} \bar{Z}_{c d}],
\end{equation}
where the second equality comes from \eqref{ZZZgZeps} (see App.~\ref{app:SS}). In particular, the two Hodge duals of the Riemann tensor do not coincide in general for spacetimes with KY symmetry.

\subsection{Contractions of Levi-Civita tensors}

Formulae for the contraction of a Levi-Civita tensor with itself (on several indices), can be found in Eq.~(B.2.13) of \cite{Wald} and read:
\begin{subequations} \label{epsIds}
	 \begin{align}         \varepsilon^{abcd}\varepsilon_{abcd} &= - 24 ,\\
    	 \varepsilon^{abch}\varepsilon_{abcd} &= - 6\, \delta^h_d , \\
    	 \varepsilon^{abgh}\varepsilon_{abcd} &= - 4\, \delta^{[g}_c \delta^{h]}_d ,\\
    	 \varepsilon^{afgh}\varepsilon_{abcd} &= - 6\, \delta^{[f}_b \delta^g_c \delta^{h]}_d , \\
    	 \varepsilon^{efgh}\varepsilon_{abcd} &= - 24\, \delta^{[e}_a \delta^f_b \delta^g_c \delta^{h]}_d .
    \end{align}
\end{subequations}

\section{Identities for the bivector basis} 
\label{app:SS}

\subsection{Basis element contractions}

Below we list several contractions on one index of the basis elements introduced in Sec. \ref{sec:biv}. These formulae simply are obtained from the definitions \eqref{defXYZ}, \eqref{symbasis1}, \eqref{symbasis2} in terms of the null tetrad, and the orthogonality properties \eqref{prodXYZ} and \eqref{orthosym}.

First, we give the products (on the inner indices) of tensors from $(Z_{ab},\bar{Z}_{ab},\mcZ_{ab})$ with those from $(X_{ab}, Y_{ab}, Z_{ab}, \mcX_{ab}, \mcY_{ab}, \mcZ_{ab})$: 
\begin{align*}
  Z_a^{\ph e} X_{e c} &= - X_{ac},\quad
  Z_a^{\ph e} Y_{e c} = Y_{ac},\quad
  Z_a^{\ph e} Z_{ec} = g_{ac},\\
  Z_a^{\ph b} \mcX_{b c} &= - \mcX_{a c},\quad
  Z_a^{\ph b} \mcY_{b c} = \mcY_{a c},\quad
  Z_a^{\ph b} \mcZ_{b c} = \bar{Z}_{a c},\\
  \bar{Z}_a^{\ph b} \mcX_{b c} &= X_{ac},\quad
  \bar{Z}_a^{\ph b} \mcY_{b c} = Y_{ac},\quad
  \bar{Z}_a^{\ph b} \mcZ_{b c} = Z_{ac}, \\
  \mcZ_a^{\ph e} X_{e c} &= - \mcX_{a c},\quad
  \mcZ_a^{\ph e} Y_{e c} = \mcY_{a c},\quad
  \mcZ_a^{\ph e} Z_{e c} = \bar{Z}_{a c},\\
  \mcZ_a^{\ph e} \mcX_{e c} &= - X_{a c},\quad
  \mcZ_a^{\ph e} \mcY_{e c} = Y_{a c},\quad
  \mcZ_a^{\ph e} \mcZ_{e c} = g_{a c}.
\end{align*}
Second, products of the symmetric tensors with $Z_{ab}$: 
\begin{align*}
  \mcL_{ab} Z^b_{\ph c} & = \mcL_{a c},\quad
  \mcN_{ab} Z^b_{\ph c} = - \mcN_{a c},\quad
  \mcZ_{ab} Z^b_{\ph c} = \bar{Z}_{a c},\\
  \mcX_{ab} Z^b_{\ph c} & = \mcX_{a c},\quad
  \mcY_{ab} Z^b_{\ph c} = - \mcY_{a c},\quad
  \mcM_{ab} Z^b_{\ph c} = \mcM_{a c}. 
\end{align*}
Third, some identities expressing $(Z_{ab},\bar{Z}_{ab},g_{ab},\mcZ_{ab})$ in terms of different combinations of $(X_{ab},Y_{ab},\mcX_{ab},\mcY_{ab})$:
\begin{align*}
  Y_d^{\ph a} X_{ac} &- X_d^{\ph a} Y_{ac} = Z_{dc},\quad
  Y_e^{\ph a} X_{ac} + X_e^{\ph a} Y_{ac} =g_{ec},\\
  \mcY_d^{\ph a} X_{ac} &- \mcX_d^{\ph a} Y_{ac} =\mcZ_{dc},\quad
  \mcY_e^{\ph a} X_{ac} + \mcX_e^{\ph a} Y_{ac}=\bar{Z}_{ec}.
\end{align*}

\subsection{Decompositions of $g_{a[c}g_{d]b}$ and $\varepsilon_{abcd}$}

Of particular interest for us is the decomposition of the following geometrical tensors
\begin{equation} \label{zzxxyygg}
  Z_{ab} Z_{cd} + 2 Y_{ab} X_{cd} + 2 X_{a b} Y_{cd} = -2 g_{a[c} g_{d]b} + \ui \varepsilon_{abcd}.
\end{equation}
This identity can be directly checked, or can be constructed from scratch using the expressions in terms of the usual expressions of $g_{ab}$ and $\varepsilon_{abcd}$ in a null tetrad. Regardless, from \eqref{zzxxyygg}, one can extract the real and imaginary parts to get the bivector decomposition of the individual, real-valued tensors $g_{a[c} g_{d]b}$ and  $\varepsilon_{abcd}$ as
\begin{subequations}\label{ggXYZ_and_vepsXYZ}
    \begin{align} 
        g_{a[c}g_{d]b} &=- \frac{1}{2} \text{Re}[Z_{ab} Z_{cd} + 2 Y_{ab} X_{cd} + 2 X_{a b} Y_{cd}], \label{ggXYZ}\\
        \varepsilon_{abcd} &= \text{Im}[Z_{ab} Z_{cd} + 2 Y_{ab} X_{cd} + 2 X_{a b} Y_{cd}]. \label{epsXYZ}
    \end{align}
\end{subequations}

The proof follows either from a direct check using \eqref{prodXYZ}, or by writing a generic decomposition on the bivector basis \eqref{defXYZ} and computing the coefficients by contraction with the basis elements, using \eqref{prodXYZ}. Note, also, that \eqref{ggXYZ} can be obtained from \eqref{epsXYZ} with \eqref{epsIds}.

\subsection{Identity for $\text{Re}[Z\bar{Z}]$} 
\label{app:ReZZbar}

We start from two copies of \eqref{zzxxyygg} with different indices
\begin{align}
  Z_{eb} Z_{ac} + 2 Y_{eb} X_{ac} + 2 X_{eb} Y_{ac} & = g_{ec} g_{ab} - g_{ea} g_{bc} + \ui \varepsilon_{ebac},\\
  Z_{ab} Z_{cd} + 2 Y_{ab} X_{cd} + 2 X_{ab} Y_{cd} & = - 2 g_{a[c} g_{d]b} + \ui \varepsilon_{abcd}.
\end{align}
Contracting with $Z_d^{\ph e}$ or $Z_d^{\ph a}$, using the
contraction formulae \eqref{epsIds}, and renaming indices produces the following two equations 
(cf.\eqref{symbasis1})
\begin{align}
  \bar{Z}_{a b} Z_{d c} + 2 Y_{a b} X_{d c} - 2 X_{a b} Y_{d c} & = Z_{a c} g_{d b} - Z_{d a} g_{b c} + \ui \mcZ_a^{\ph e} \varepsilon_{e b d c},\\
  Z_{c d} \bar{Z}_{b a} - 2 Y_{c d} X_{a b} + 2 X_{c d} Y_{a b} & = g_{c a} Z_{b d} - Z_{b c} g_{d a} + \ui \mcZ_b^{\ph e} \varepsilon_{c d e a}.
\end{align}
Now taking the difference and using the symmetries of involved quantities readily leads to 
\begin{equation} \label{ZZZgZeps}
  \bar{Z}_{ab} Z_{cd} =-\mcZ_{a[c} g_{d]b} - g_{a[c} \mcZ_{d]b} - \ui \mcZ_{[a}^{\ph e} \varepsilon_{b]ecd}.
\end{equation}
Finally, taking the real and imaginary parts and recalling that $\mcZ_{ab}$  is real-valued leads to the following identities which are central to the results of Sec.~\ref{sec:PBint}:
\begin{equation}\label{ReZZbar}
  \text{Re} [\bar{Z}_{a b} Z_{c d}] = -\mcZ_{a [c} g_{d] b} - g_{a [c} \mcZ_{ d] b} \quand
  \text{Im} [\bar{Z}_{a b} Z_{c d}] = -\mcZ_{[a}^{\ph e}
  \varepsilon_{b] e c d}.
\end{equation}

\section{Proof of identity \eqref{eq:C_Psi}}
\label{app:C_Psi}

This appendix shows that $\Psi_2 C^3$ in~\eqref{eq:C_Psi} is covariantly constant in 4D Einstein ($\Lambda$-vacuum) spacetimes admitting a rank-4 KY tensor.


Consider the following expressions for the Riemann tensor of a type-D, Einstein spacetime, as obtained by the Weyl decomposition \eqref{Weyldec} and the bivector decompositions \eqref{CZZggeps}:
\begin{equation} \label{zipzap}
  R_{abcd} = \text{Re} \left[ 3 \Psi_2 Z_{a b} Z_{c d} + \left( \Psi_2 +
  \Lambda/3 \right) (2 g_{a [c } g_{ d] b} -
  \ui \varepsilon_{a b c d}) \right].
\end{equation}
We also consider the self-dual tensor $\mcR_{a b c d} = R_{abcd} - \ui R^{\star}_{a b c
d}$, so that its Hodge dual satisfies $\mcR_{a b c d}^{\star} = \ui \mcR_{a b c d}$. 
Since $R_{abcd}$ satisfies the differential Bianchi identity, 
it follows that the Hodge dual $R^{\star}_{a b c d}$ does too: 
$\nabla_{[a } R^{\star}_{b c] d e} = \tfrac{1}{2} \nabla_{[a } R_{b  c] f
g} \varepsilon^{f g}_{\phantom{a b} d e} = 0^{}$. 
Therefore $\mcR_{a b c d}$ satisfies the Bianchi identity as well. Following \eqref{zipzap}, it admits the following bivector decomposition
\begin{equation}\label{c-zipzap}
  \mcR_{a b c d}  = 3 \Psi_2 Z_{a b} Z_{c d} + \left( \Psi_2 +
  \Lambda/3 \right) (2 g_{a [c } g_{ d] b} -
  \ui \varepsilon_{a b c d}).
\end{equation}

We next take a covariant derivative $\nabla_e$ of $\mcR_{a b c d}$ in \eqref{c-zipzap}, and contract with $Z^{c d}$. Using $Z^{c d} Z_{c d} = - 4$, $Z^{c d} \nabla_e Z_{c d} = 0$, $Z^{\star}_{a b} = \ui Z_{a b}$ and $\nabla_a g_{bc} = \nabla_a \varepsilon_{bcde} = \nabla_a \Lambda = 0$, we have
\begin{equation}\label{eq:nabla_mcR}
  \nabla_e (\mcR_{a b c d}) Z^{c d}  =  - 4 (2 \nabla_e \Psi_2 Z_{a b} + 3 \Psi_2 \nabla_e Z_{a b}).
\end{equation}
We notice that at this step that the dependence on $\Lambda$ is lost. Now we antisymmetrize over indices $e\, a\, b$ on each side in \eqref{eq:nabla_mcR}, and use the Bianchi identity for the LHS. We are left with 
\begin{equation}
  0 = 2 \nabla_e \Psi_2 Z_{a b} + 3 \Psi_2 \nabla_e Z_{a b} + 2 \nabla_a
  \Psi_2 Z_{b e} + 3 \Psi_2 \nabla_a Z_{b e} + 2 \nabla_b \Psi_2 Z_{e a} + 3
  \Psi_2 \nabla_b Z_{e a} .
\end{equation}
Contracting this with $Z^{a b}$ and using $Z^{a b} Z_{b e} = \delta^a_e$ leads to
\begin{equation}\label{eq:id-Psi2}
  2 \nabla_e \Psi_2 = 3 \Psi_2 Z^{a b} \nabla_a Z_{b e} .
\end{equation}

Lastly, we can get an expression for the term $Z^{a b} \nabla_a Z_{b e}$ in \eqref{eq:id-Psi2}, in terms of the KY properties. Consider the self-dual, complex quantity $\Phi_{a b} = f_{a b} - \ui f_{a b}^{\star}$ (cf. \eqref{defPhi}). Using Eqs.~\eqref{defxi} and \eqref{dfstar}, we find that $\Phi_{ab}$ satisfies 
\begin{equation} \label{CCKY}
  2 \nabla_{(a } \Phi_{ b) e} =  g_{a b} \Phi_e - \Phi_{(a } g_{ b) e}, \quad \text{with} \quad 3 \Phi^a := \nabla_c \Phi^{c a} .
\end{equation}
which makes $\Phi_{ab}$ a complex conformal Killing-Yano tensor \cite{JeziLuka.06}. 
Combining this result with the KY decomposition \eqref{defPhi} gives $C Z^{a e} \nabla_a Z_{b e}  =  -2 \nabla_b C$. Along with \eqref{eq:id-Psi2}, this finally leads to
\begin{equation}\label{Psi2C3cst}
  \nabla_a (\Psi_2 C^3) = 0,
\end{equation}
which is the result \eqref{eq:C_Psi}. As a consistency check, from Eqs.~(3.9)-(3.10) of \cite{Ha.20}, we have $\Psi_2 C^3 = \ui M$ in Kerr, where $M$ is the Kerr mass parameter.

\section{Symmetries of $F_{abcd}$ and related identities.} 
\label{app:Fa(bc)d}

\subsection{Proof of identity~\eqref{Fa(bc)d=0}}

In this appendix, our goal is to show the identity~\eqref{Fa(bc)d=0}, i.e., 
\begin{equation}\label{eq:def-tildeF}
    \tilde{F}_{a(bc)d} = 0, \quad \text{where} \quad 
    \tilde{F}_{abcd} := F_{abcd} + 4\tilde{M}_{abcd}.
\end{equation}
To show this algebraic symmetry without complicated nested symmetry operators, we will instead show the equivalent identity 
\begin{equation}
    \tilde{F}_{abcd} w^b w^c=0, 
\end{equation}
for an arbitrary vector $w^a$. 
Below we give the proof under the Ricci-flat assumption $R_{ab} = 0$. Since all formulae are linear-in-$R_{abcd}$, the extension to Einstein spacetimes amounts to verifying the same identity with 
$\tfrac{2}{3} \Lambda g_{a[c}g_{b]d}$, which is straightforward.

We start by contracting the definition of $\tilde{F}_{abcd}$ (cf. \eqref{Mtildefinal}-\eqref{Fabcd}) with $w^bw^c$. 
The contributions from the first and the second terms in \eqref{Fabcd} are 
\begin{subequations} \label{lines}
    \begin{align}
  \tilde{F}_{abcd} w^b w^c &= 
  \left(R_{abde}K^e_{\ph c}+ \tfrac{2}{3} \nabla_d \nabla_{[a } K_{b]c} + 4 \tilde{M}_{a b c d}  \right) w^b w^c, \label{line0}\\ 
  &= \left( \tfrac{1}{2} R_{a b e f} f^{e f} Y_{c d} + \mathfrak{Z} R_{a b c d}^{\star} +  f_a^{\ph f} R_{f b d e} f^e_{\ph c} - f_b^{\ph f} R_{fa d e} f^e_{\ph c} \right. \label{line1} \\
  & + \left. 2 f_{e [a}^{\star} g_{ b] c} \nabla_d \xi^e - f_{a b}^{\star} \nabla_c \xi_d - \tfrac{1}{4} (f_{a b} R_{c d f e} f^{f e} + f_{c d} R_{a b f e} f^{f e}) - \mathfrak{Z} R_{c d a b}^{\star} \right) w^b w^c, \label{line2}
  \end{align}
\end{subequations}
where: the quadratic-in-$\xi^a$ terms from $\tilde{M}_{abcd}$ [cf. \eqref{Mtildefinal}] and from the second covariant derivative of $K_{ab}$ [cf. \eqref{KSdef} and \eqref{defxi}] have cancelled each other; and the first line \eqref{line1} corresponds to the term $R_{abde}K^e_{\ph c}$ in \eqref{line0}, obtained by combining equations \eqref{intcondcentral} and \eqref{fffeps}.
The contribution from the third term in \eqref{Fabcd} vanishes identically since $\varepsilon_{abcd}$ is totally antisymmetric. 

Next, we note that the two Hodge duals of the Riemann tensor annihilate (cf. \eqref{duals} with $\phi_2=0$ for Einstein spacetimes), and we re-write the first term in line \eqref{line2}  by combining equations \eqref{Dxi} and \eqref{epsIds}. The latter operation brings a term that results, crucially, in a change of sign $+\rightarrow-$ within the brackets of line \eqref{line2}. Ultimately, \eqref{lines} becomes
\begin{align}\label{ploup}
  \tilde{F}_{abcd} w^b w^c &= 
  \left(\frac{1}{4} g_{c [a } g_{ b] d} R_{gh ef}f^{gh} f^{ef} - \tfrac{1}{2} f^{ef} R_{ef d g}g_{c [a } f_{ b]}^{\ph g} - f_{a b}^{\star} \nabla_c \xi_d \right. \notag \\
  & \left.\qquad - \frac{1}{4} (f_{a b} R_{c d f g} f^{f g} - f_{c d} R_{a b f g} f^{f g}) - 2 f_{[a }^{\ph  f} R_{ b] f d e} f^e_{\ph c} \right) w^b w^c.
\end{align}
It is then just a matter of replacing these 5 terms with their bivector decompositions:
\begin{subequations}
    \begin{align}
    \nabla_c \xi_d & =  - \text{Im} [\Psi_2 C Z_{c d}], \\ 
    R_{ghef} f^{gh} f^{ef} & =  8 \text{Re} [\Psi_2 C^2], \\
    f^{e f} R_{e f d g} g_{c [a } f^{\phantom{c} g}_{ b]} &= 2 \text{Re} [\Psi_2 (C^2 g_{c [a } g_{b] d} + | C|^2 g_{c [a } Z_{ b] d})], \\
    f^{\ph  f}_{[a } R_{ b] f d e} f^e_{\ph c} & =  \tfrac{1}{2} \text{Re} [\Psi_2 C (2 Z_{d [a }f_{ b] c} + C g_{c [a } g_{ b] d} + \bar{C} Z_{c [a } g_{ b] d})], \\
    f_{a b} R_{c d f g} f^{f g} & =  f_{c d} R_{a b f g} f^{f g} + 4 \text{Im}[\Psi_2] | C |^2 \text{Im} [\bar{Z}_{a b} Z_{c d}],
    \end{align}
\end{subequations}
obtained by combining \eqref{fNP},  \eqref{fstarNP} and \eqref{CXYZ} with the bivector products in App.~\ref{app:SS}. Once substituted into \eqref{ploup}, the terms $\propto g_{c[a}g_{b]d}$ cancel, leaving
\begin{align}
        \tilde{F}_{abcd}w^bw^c &= -|C|^2 \Bigl(\operatorname{Re} [\Psi_2] (g_{c [a } \mathcal{Z}_{ b] d} + \mathcal{Z}_{c[a } g_{ b] d}) + \operatorname{Im} [\Psi_2] \operatorname{Im} [\bar{Z}_{a b} Z_{c d}]\Bigr) w^b w^c \notag  \\
        & \qquad - \left(2 \operatorname{Re} [\Psi_2 C Z_{ d [a }]\, f_{ b] c} - \operatorname{Im} [C Z_{a b}] \,\operatorname{Im} [\Psi_2 C Z_{c d}] \right) w^b w^c.
\end{align}
In the first line, we use identity \eqref{ReZZbar}. In the second line, we expand the anti-symmetry of the first term: one term vanishes with $w^b w^c$, and for the other one we use \eqref{fNP} one last time. We finally obtain
\begin{align} \label{poupidou}
    \tilde{F}_{abcd} w^b w^c &= |C|^2 \left( \operatorname{Re} [\Psi_2] \operatorname{Re} [\bar{Z}_{a b} Z_{c d}] - \operatorname{Im}[\Psi_2] \operatorname{Im} [\bar{Z}_{a b} Z_{c d}] \right) w^b w^c \cr
    & \qquad - \left(\operatorname{Re} [\Psi_2 C Z_{c d} C Z_{a b}] + \operatorname{Im} [C Z_{a b}] \operatorname{Im} [\Psi_2 C Z_{c d}] \right) w^b w^c.
\end{align}
Using the elementary identities 
$\text{Re}[zz']=\text{Re}[z]\text{Re}[z']-\text{Im}[z]\text{Im}[z']$, $\text{Re}[z]=\text{Re}[\bar{z}]$ and $\text{Im}[z]=-\text{Im}[\bar{z}]$, valid for any $(z,z') \in \CC^2$, one finds that, upon expanding in real and imaginary parts, all contributions on the RHS of \eqref{poupidou} cancel term by term. Hence the right-hand side vanishes identically, we have shown that $\tilde{F}_{a(bc)d}=0$. 

\subsection{Proof of Eqs.~\eqref{FMeps} and \eqref{DZ}}

Above we have shown that $\tilde F_{abcd}$ in \eqref{eq:def-tildeF} has the symmetries that is antisymmetric in $bc$, but it is also anti-symmetric in $ab$ and $cd$. Therefore, it is completely antisymmetric, and there exist a scalar field $\mathfrak{F}$ such that 
\begin{equation}
    \tilde F_{abcd} = \mathfrak{F} \, \varepsilon_{abcd} \quad \Rightarrow \quad \mathfrak{F} = -\frac{1}{4!} \tilde F_{abcd}\, \varepsilon^{abcd},
\end{equation}
where the second equality follows from \eqref{epsIds}. Let us now compute $\mathfrak{F}$ by using the above equality and contracting the definition of $F_{abcd}$ and $\tilde{M}_{abcd}$, cf. \eqref{Fabcd} and \eqref{Mtildefinal}, respectively, with the Levi-Civita tensor $\varepsilon^{abcd}$. When doing so, the first and second terms in \eqref{Fabcd} vanish by symmetry of $K_{ab}$ and the Bianchi identity $R_{[abc]d}=0$. Similarly, the first and second terms in in \eqref{Mtildefinal} vanish by symmetry of the metric. Combining all this leads to
\begin{subequations}
    \begin{align}
        \tilde F_{abcd}\, \varepsilon^{a b c d}        
        &=\frac{4}{3} \varepsilon^{a b c d} \varepsilon_{a b c e} \nabla_d \nabla^e \mathfrak{Z} - \frac{1}{2} f_{c d} R_{a b ef} f^{ef} \varepsilon^{a b c d} - \mathfrak{Z} R_{a b c d}^{\star} \varepsilon^{a b c d}, \\
        &= - \,8\, \Delta \mathfrak{Z} - R_{a b c d} f^{a b}_{\star} f^{c d} + 8 \Lambda \mathfrak{Z} \label{bipoup}
    \end{align}
\end{subequations}
where we applied the Levi-Civita identities in App. \ref{epsIds} to the first and third term and used the definition of the Hodge dual to the second term. Then, equation \eqref{bipoup} can be simplified by combining \eqref{deff}, \eqref{dfstar} and \eqref{Dxi} to obtain the following identity
\begin{equation}
    \frac{1}{4} R_{a b c d} f^{a b} f^{c d}_{\star} = \Delta \mathfrak{Z} + 2 \Lambda \mathfrak{Z}.
\end{equation}
The above equation, when inserted into \eqref{bipoup} readily gives $\mathfrak{F}=\tfrac{1}{2}\Delta\mathfrak{Z}$ (as claimed in \eqref{FMeps}), and when combined with the Weyl decomposition \eqref{Weyldec} gives the claimed identity \eqref{DZ}.

\section{Covariant proof of Eqs.~(103) of Ref.~\cite{ComDruVin.23}} 
\label{app:BLtoCov}

First, we note that equations (103) of \cite{ComDruVin.23} can be simplified under the TD SSC, by setting the quantity $\mathcal{A}$ there to $0$, cf. Eq. (10) there. We also note that the third equation in (103) is a mere consequence of the second one. Therefore, using their definitions (80), all we have to show is that (in the notations of \cite{ComDruVin.23}):
\begin{align}
  (2 N_{ac}  \bar{N}_{bd} - g_{cd} h_{ab} - g_{ab} h_{cd}) s^a s^b  {\hat p}^c {\hat p}^d & = 0,\\
  (h_{a c} N_{d b} + h_{a b} N_{c d} + h_{d b} N_{a c} + g_{a b}  \bar{N}_{c d} + g_{d b} \bar{N}_{a c}) s^a s^b {\hat p}^c \xi^d & = 0,
\end{align}
where $s^a:=S^a/S_\circ$ is a unit vector (such that $s^a s_a=1$), and ${\hat p}^a$ is the unit momenta~\eqref{eq:def-hatp}. 
In our notations, using the dictionary \ref{table:dictionnary} these identities read
\begin{subequations}\label{toprov}
\begin{align}
  \left( 2 Z_{c (a} \bar{Z}_{ b) d} - Z_{a b} g_{c d}  - g_{a b} Z_{c d} \right) s^a s^b {\hat p}^c {\hat p}^d & = 0, \\ 
  \left( \mcZ_{a b} Z_{c d} + \mcZ_{a c} Z_{d b} + Z_{a c} \mcZ_{d b} - g_{a b} \bar{Z}_{c d} - g_{d b} \bar{Z}_{a c} \right) s^a s^b {\hat p}^c \xi^d &=0.
\end{align}
\end{subequations} 

To prove these relations, we introduce $L_{ab}=2\ell_a n_b\in\RR$ and $M_{ab}=2m_a\bar{m}_b\in\ui\RR$. These tensors have the symmetric property $L_{ab} L_{cd} = L_{ad} L_{cb}$, and similarly for $M_{ab}$. In addition, their symmetric and anti-symmetric parts can be used to construct all the geometric tensors involved in \eqref{toprov}, as follows:
\begin{equation} \label{LandM}
  Z_{ab} = L_{[ab]} - M_{[ab]},\quad
  \bar{Z}_{ab} = L_{[ab]} + M_{[ab]},\quad
  g_{ab} = M_{(ab)} - L_{(ab)},\quad
  \mcZ_{ab} = -L_{(ab)} - M_{(ab)}.
\end{equation}

Lastly, by definition of the spin vector \eqref{zip!}, one has ${\hat p}_a S^a=0$. Using equation \eqref{LandM}, this gives the relation $M_{(ab)} s^a {\hat p}^b = L_{(ab)} s^a {\hat p}^b$. By combining this with equations \eqref{LandM} and the aforementioned symmetries, equations \eqref{toprov} directly follow.

\end{document}